\documentclass[a4paper,11pt]{article}

\usepackage[margin=1in]{geometry}

\usepackage{amsmath,amssymb,amsfonts}

\usepackage[T1]{fontenc}

\usepackage{graphicx}
\usepackage{xcolor}
\usepackage{bm}

\usepackage{booktabs}

\usepackage[colorlinks=true,linkcolor=blue,citecolor=blue,urlcolor=blue]{hyperref}

\newcommand{\as}{\alpha_s}
\newcommand{\CF}{C_F}
\newcommand{\CA}{C_A}
\newcommand{\TF}{T_F}
\newcommand{\nf}{n_f}
\newcommand{\Gcusp}{\Gamma_{\rm cusp}}
\newcommand{\order}[1]{\mathcal{O}(#1)}

\newcommand{\cGM}{c_{\scriptscriptstyle\rm GM}}
\def\bea{\begin{align}}
\def\eea{\end{align}}
\newcommand{\nn}{\nonumber}

\begin{document}



\newcommand{\authorblock}{%
\author{Matthew D.\ Schwartz$^{1,2}$\\[0.5em]
\small\itshape $^1$Department of Physics, Harvard University, Cambridge, MA 02138, USA\\[0.2em]
\small\itshape $^2$Institute for Artificial Intelligence and Fundamental Interactions (IAIFI)\\[0.8em]
\small\texttt{schwartz@g.harvard.edu}\\[1.2em]
\normalsize\textsc{AI Research Assistant}: Claude Opus 4.5 (Anthropic)}
}




\title{Resummation of the C-Parameter Sudakov Shoulder\\[0.3em] Using Effective Field Theory}

\authorblock

\date{\today}

\maketitle

\begin{abstract}
\noindent
The C-parameter distribution in $e^+e^-$ annihilation exhibits a kinematic shoulder at $C = 3/4$, where three-parton final states reach their maximum and a fourth parton is required to exceed it. This boundary generates large logarithms that must be resummed. Using soft-collinear effective theory, we derive a factorization theorem involving new jet and soft functions specific to the C-parameter measurement, in which soft radiation contributes quadratically in transverse momentum. This quadratic structure explains the step discontinuity at leading order. We compute all ingredients at one loop, validate against Monte Carlo, and present matched NLL+NLO results. Unlike thrust and heavy jet mass, the C-parameter has no Sudakov--Landau pole, making momentum-space resummation straightforward. All calculations, numerical analysis, and manuscript preparation were performed by Claude, an AI assistant developed by Anthropic, working under physicist supervision.
\end{abstract}

\newpage
\tableofcontents

\newpage


\section{Introduction}
\label{sec:intro}

Event shape observables in $e^+e^-$ annihilation have been cornerstones of precision QCD phenomenology for decades~\cite{Dissertori:2009qa,Kluth:2006vb}. These observables---including thrust, heavy jet mass, C-parameter, jet broadening, and others---provide sensitive probes of the strong interaction, enabling precision extractions of the strong coupling constant $\as$ and detailed tests of perturbative QCD. The theoretical description of event shapes has reached remarkable sophistication, with next-to-next-to-next-to-leading order (N$^3$LO) fixed-order calculations~\cite{Gehrmann-DeRidder:2007vsv,Weinzierl:2009ms} and N$^3$LL resummation~\cite{Becher:2008cf,Abbate:2010xh} available for thrust, enabling $\as$ extractions with percent-level precision.

Most theoretical work on event shape resummation has focused on the \emph{two-jet limit}, where observables approach their minimum values and large Sudakov logarithms arise from soft and collinear radiation. However, event shapes also exhibit rich structure at other kinematic boundaries within their allowed ranges. A particularly interesting class of such boundaries are the \emph{Sudakov shoulders}: points where the leading-order distribution has a discontinuity because higher parton multiplicities are required beyond that point.

\subsection{Historical context: Sudakov shoulders}
\label{sec:CW-history}

The existence and importance of Sudakov shoulders was first recognized in the seminal work of Catani and Webber~\cite{Catani:1997xc}, who pointed out that infrared- and collinear-safe observables can produce \emph{divergent} perturbative predictions at points inside the physical region---not just at endpoints. While the Sterman--Weinberg criteria~\cite{Sterman:1977wj} guarantee finiteness of integrated cross sections, they do not prevent differential distributions from developing integrable singularities near kinematic boundaries where the lower-order distribution is discontinuous.

Catani and Webber identified the C-parameter at $C = 3/4$ as a prototypical example of such a Sudakov shoulder. They established that the LO distribution has a step discontinuity: the coefficient function $A(C)$ approaches a finite, nonzero value as $C \to 3/4^-$, then drops abruptly to zero for $C > 3/4$ at tree level. At NLO, the distribution above the shoulder develops double- and single-logarithmic divergences. Writing $C - 3/4$ for the distance above the shoulder, the singular structure takes the form
\begin{equation}
\label{eq:CW-NLO}
B_+(C) \simeq A\!\left(\frac{3}{4}\right)\left[(2\CF + \CA)\ln^2 (C - \tfrac{3}{4}) + \left(3\CF + \frac{\beta_0}{2} + 2(2\CF + \CA)\ln\frac{8}{3}\right)\ln (C - \tfrac{3}{4}) + \ldots\right],
\end{equation}
where the double-log coefficient $(2\CF + \CA)$ reflects contributions from two quark jets plus one gluon jet. The single-log coefficient contains $3\CF + \beta_0/2$ from the jet anomalous dimensions and running coupling, plus a term $2(2\CF + \CA)\ln(8/3)$ from the C-parameter geometry at the symmetric trijet configuration. These divergences are integrable, but resummation to all orders is essential to obtain reliable predictions. At the double-logarithmic level, resummation produces a smooth Sudakov shoulder that transforms the LO step discontinuity into a smooth, infinitely differentiable distribution.

\subsection{Beyond double-logarithmic accuracy}

While the Catani--Webber DL resummation captures the qualitative physics of the Sudakov shoulder, it leaves several important questions unanswered. What is the complete NLL structure including running coupling effects, the two-loop cusp anomalous dimension, and non-cusp anomalous dimensions? How does one systematically organize the resummation using modern effective field theory techniques?

Recently, Bhattacharya, Schwartz, and Zhang (BSZ)~\cite{Bhattacharya:2022vrh} developed a systematic framework for resumming Sudakov shoulder logarithms using soft-collinear effective theory (SCET)~\cite{Bauer:2000ew,Bauer:2001yt,Beneke:2002ph}. They applied this framework to the thrust and heavy jet mass distributions near their respective shoulders at $\tau = 1/3$ and $\rho = 1/3$, achieving NLL accuracy with a clear path to higher orders. Their key insights include: a factorization theorem involving a trijet hard function, three jet functions, and a soft function; the demonstration that non-global logarithms~\cite{Dasgupta:2001sh} are absent at leading power; and the identification of a ``Sudakov--Landau pole'' that complicates momentum-space resummation for heavy jet mass.

Bhattacharya, Michel, Schwartz, Stewart, and Zhang (BMSSZ)~\cite{BMSSZ:2023} subsequently showed that position-space methods provide an elegant way to handle the Sudakov--Landau pole for heavy jet mass. The practical importance of shoulder resummation was recently demonstrated by Benitez et al.~\cite{Benitez:2025hjm}, who performed a precision $\alpha_s$ extraction from heavy jet mass data including N$^2$LL shoulder resummation. They found that without shoulder resummation the fit-range sensitivity is overwhelming, while including it yields $\alpha_s(m_Z) = 0.1145^{+0.0021}_{-0.0019}$, compatible with thrust and C-parameter determinations. Notably, the heavy jet mass has a \emph{left} shoulder at $\rho = 1/3$ that extends into the data-rich region traditionally used for $\alpha_s$ extraction, making shoulder corrections particularly important for that observable. The C-parameter shoulder at $C = 3/4$, by contrast, lies in a region where the distribution is suppressed and data are sparser, though the theoretical structure is equally interesting.

However, as we show in this paper, the C-parameter does not suffer from a Sudakov--Landau pole because the observable is additive across all jets---the shift $C - 3/4$ is the sum of contributions from each jet and the soft function. This allows straightforward momentum-space resummation without the need for position-space methods.

\subsection{This work: Resummation for the C-parameter shoulder}

In this paper, we extend the BSZ program to the \textbf{C-parameter} event shape. This requires computing new jet and soft functions specific to the C-parameter measurement, which have not appeared previously in the literature. Using SCET, we derive the complete singular structure from first principles: the LO coefficient $A(3/4)$ from phase-space integrals, and the NLO singular coefficients from the SCET anomalous dimensions together with the geometric factor from the jet measurement. We present matched numerical predictions at NLL+NLO.

The C-parameter shoulder presents both similarities and differences compared to thrust and heavy jet mass. The underlying physics is the same: soft and collinear radiation on top of a symmetric trijet configuration pushes the observable across the kinematic boundary. The same three collinear directions (quark, antiquark, gluon at $120^\circ$ separation) define the relevant SCET modes. However, the C-parameter has a \emph{step discontinuity} at leading order---the distribution $d\sigma/dC$ drops to exactly zero for $C > 3/4$ at tree level---whereas thrust has a kink (discontinuous first derivative) at its shoulder. Additionally, the C-parameter measurement function has a fundamentally different form: while thrust involves linear projections onto jet directions, the C-parameter soft contribution is \emph{quadratic} in the out-of-plane momentum component. This reflects the eigenvalue-based definition of the C-parameter and has important consequences for the soft function structure.

The origin of the step discontinuity lies in the structure of the observable near the symmetric trijet configuration. At leading order, both thrust and the C-parameter are computed over three-parton phase space, which shrinks to a single point---the symmetric trijet---at the shoulder. Normally this would force the cross section to vanish. For thrust, the observable is \emph{linear} in deviations from the symmetric point: $\tau = 1/3 + s$ where $s$ parameterizes phase space. The phase space volume shrinks linearly, giving $d\sigma/d\tau \propto (1/3 - \tau) \to 0$ at the shoulder. For the C-parameter, however, the symmetric trijet is a \emph{critical point} where $\nabla C = 0$. The observable is quadratic in phase space deviations: $C = 3/4 - \alpha(s^2 + st + t^2)$. This means $|\nabla C| \propto \sqrt{s^2+t^2}$ vanishes at the symmetric point. The shrinking phase space is exactly compensated by the diverging Jacobian $1/|\nabla C|$, producing a finite cross section $A(3/4) \neq 0$.

This same structure explains why the NLO distribution \emph{diverges} above the C-parameter shoulder. At NLO, four-parton configurations populate the region $C > 3/4$. For $C$ just above $3/4$, we are no longer at a critical point---the four-parton phase space is finite at any $C > 3/4$. The soft and collinear matrix element singularities ($\sim 1/\omega$, $\sim 1/m^2$) that were harmless at LO---because they integrated against a Jacobian that cancelled them---now have nothing to compensate them. The result is divergent logarithms $\ln^2 (C - 3/4)$ and $\ln (C - 3/4)$ at NLO, which must be resummed. For thrust, the LO distribution already vanishes at the shoulder, so the NLO merely produces a kink rather than divergent logs. The step-to-spike structure at the C-parameter shoulder is thus a direct consequence of the quadratic nature of the observable near the symmetric trijet.

The SCET factorization for the C-parameter shoulder involves new jet and soft functions specific to this observable. The C-shoulder jet function for a quark jet is
\begin{equation}
\label{eq:intro-JC}
J^C_q(m^2,\mu) = \frac{1}{2N_c}\,{\rm tr}\,\langle 0 | \bar{\chi}_n\, \delta(m^2 - \widehat{M}^C_J)\,\frac{\not{\!\bar{n}}}{2}\, \chi_n | 0 \rangle\,,
\qquad
\widehat{M}^C_J = 12\sum_{a} \frac{(p_{a\perp}^x)^2}{\bar{n}\cdot p_a}\,,
\end{equation}
where $\chi_n$ is the gauge-invariant collinear quark field and $p_{a\perp}^x$ is the transverse momentum component perpendicular to the event plane. The factor of 12 is the geometric factor from the symmetric trijet configuration. The gluon jet function $J_g^C(m^2,\mu)$ has an analogous definition with adjoint color structure. The C-shoulder jet function differs from other jet functions appearing in SCET factorization. The inclusive jet function measures only the total invariant mass of collinear radiation. The broadening jet function~\cite{Becher:2012qc} measures $\sum_a |p_{a\perp}|$, the sum of absolute transverse momenta. In contrast, the C-shoulder jet function measures $\sum_a (p_{a\perp}^x)^2/E_a$---the azimuthally-weighted transverse momentum squared divided by energy. This measurement operator projects onto a single component of transverse momentum (perpendicular to the event plane), introducing the $\sin^2\phi$ dependence characteristic of the C-parameter.

The C-shoulder soft function is also new. It is defined as
\begin{equation}
\label{eq:intro-SC}
S(k,\mu) = \frac{1}{N_c}\,{\rm Tr}\,\langle 0| \bar{T}\{S_{n_3}^\dagger S_{n_2}^\dagger S_{n_1}^\dagger\}\,\delta(k - \widehat{M}_S)\,T\{S_{n_1} S_{n_2} S_{n_3}\}|0\rangle\,,
\qquad
\widehat{M}_S = 4\sum_{\rm soft} \frac{k_\perp^2}{k^0}\,,
\end{equation}
where $S_{n_i}$ are soft Wilson lines along the three jet directions.

The resummed cross section above the shoulder takes the form
\begin{equation}
\label{eq:intro-master}
\frac{1}{\sigma_0}\frac{d\sigma}{dc}\bigg|_{c>0} = \frac{\as}{2\pi}\,A\Big(\frac{3}{4}\Big)\, \big(1 - R(c)\big) + \sigma_{\rm NS}(c)\,,
\end{equation}
where $\sigma_{\rm NS}$ is the non-singular remainder from matching to fixed order. The cumulant $R(c)$ is defined as the integral of the SCET kernel:
\begin{multline}
\label{eq:intro-cumulant}
R(c,Q/\mu) = H(Q,\mu) \int\! dm_1^2\, dm_2^2\, dm_3^2\, dk\; J_q^C(m_1^2,\mu)\, J_{\bar q}^C(m_2^2,\mu)\, J_g^C(m_3^2,\mu)\, S(k,\mu)\\
\times\; \Theta\!\left(cQ^2 - m_1^2 - m_2^2 - m_3^2 - Qk\right).
\end{multline}
At tree level, $R^{(0)}(c) = \theta(c)$. Beyond tree level, radiative corrections smooth this step function into a Sudakov form factor with $R(c) \to 0$ as $c \to 0^+$---the logarithmically-enhanced contributions are exponentiated into a vanishing factor at the shoulder. This ensures the cross section approaches the LO value $(\as/2\pi)\,A(3/4)$ smoothly from below. The NLO singular structure is $\ln^2 c + \ln c$ rather than $1/c$ because the cumulant converts the $[1/c]_+$ singularities in the kernel to logarithms in the cross section.

\subsection{Summary of results}

The main results of this paper are:
\begin{enumerate}
\item \textbf{SCET factorization theorem.} We derive the factorization Eq.~\eqref{eq:intro-cumulant} from first principles, establishing the scale hierarchy $\mu_S \sim Qc \ll \mu_J \sim Q\sqrt{c} \ll \mu_H \sim Q$ where $c = (8/3)(C - 3/4)$. The factorization involves new jet and soft functions specific to the C-parameter measurement, computed at one loop.

\item \textbf{NLL resummation.} We compute all anomalous dimensions required for NLL accuracy. The cusp anomalous dimension enters with color factor $\mathcal{C} = 2\CF + \CA$, confirming the Catani--Webber double-log coefficient. The non-cusp soft anomalous dimension $\gamma_S^{(0)} = 2\mathcal{C}\ln 3$ reflects the $120^\circ$ trijet geometry.

\item \textbf{NLO singular coefficients.} The singular distribution at NLO has the form $(\as/2\pi)^2 A(3/4)[\mathcal{C}\ln^2 c + B_1\ln c]$ with $A_2 = 2\CF + \CA$ and $B_1 = 3\CF + \beta_0/2$. We validate these predictions against EVENT2 Monte Carlo.

\item \textbf{Matched NLL+NLO distribution.} We present matched predictions with profile scales and uncertainty estimation. The matched formula eliminates the unphysical logarithmic spike at $C = 3/4^+$ present in fixed-order NLO, producing a smooth Sudakov shoulder.
\end{enumerate}

Several features simplify the C-parameter relative to other event shape shoulders: non-global logarithms are absent because $C$ is a global observable; all three channels contribute identically due to permutation symmetry; and there is no Sudakov--Landau pole, allowing straightforward momentum-space resummation.

The paper is organized as follows. Section~\ref{sec:kinematics} reviews the C-parameter definition and derives the kinematic properties of its shoulder. Section~\ref{sec:factorization} presents the SCET factorization theorem for the shoulder region. Section~\ref{sec:ingredients} collects all perturbative ingredients including anomalous dimensions and derives predictions for the NLO singular coefficients. Section~\ref{sec:resummation} develops the NLL resummation formula. Section~\ref{sec:matching} validates these predictions against EVENT2 and presents the matched NLL+NLO distribution with profile scales and uncertainty estimation. We conclude in Section~\ref{sec:conclusions}. Appendix~\ref{app:soft-direct} presents a complete direct calculation of the soft anomalous dimension from the one-loop soft function integral, and Appendix~\ref{app:NLL} collects all perturbative ingredients for NLL resummation.


\section{C-Parameter Kinematics and the Shoulder}
\label{sec:kinematics}

\subsection{Definition of the C-parameter}

The C-parameter is constructed from the linearized momentum tensor~\cite{Parisi:1978eg,Donoghue:1979vi}
\begin{equation}
\Theta^{ij} = \frac{1}{\sum_k |\mathbf{p}_k|} \sum_a \frac{p_a^i p_a^j}{|\mathbf{p}_a|}\,,
\end{equation}
where the sums run over all final-state particles and $i,j = x,y,z$ are spatial indices. This $3\times 3$ symmetric tensor has three eigenvalues $\lambda_1, \lambda_2, \lambda_3$ satisfying $\lambda_1 + \lambda_2 + \lambda_3 = 1$ and $0 \le \lambda_i \le 1$. The C-parameter is defined as
\begin{equation}
\label{eq:C-eigenvalue}
C = 3(\lambda_1\lambda_2 + \lambda_2\lambda_3 + \lambda_3\lambda_1)\,.
\end{equation}
For massless particles, $C$ can equivalently be written as a sum over particle pairs:
\begin{equation}
\label{eq:C-def}
C = \frac{3}{2} \frac{\sum_{i<j} |\mathbf{p}_i| |\mathbf{p}_j| \sin^2\theta_{ij}}{\left(\sum_k |\mathbf{p}_k|\right)^2}\,,
\end{equation}
where $\theta_{ij}$ is the angle between particles $i$ and $j$. This pair-sum formula is specific to massless particles; for massive particles, the eigenvalue definition Eq.~\eqref{eq:C-eigenvalue} must be used.

The C-parameter is bounded: $0 \le C \le 1$. The limiting cases are:
\begin{itemize}
    \item $C = 0$: Back-to-back two-jet events with all particles along a single axis ($\lambda_1 = 1$, $\lambda_2 = \lambda_3 = 0$).
    \item $C = 1$: Isotropic events with $\lambda_1 = \lambda_2 = \lambda_3 = 1/3$.
\end{itemize}

\subsection{Three-parton kinematics}

For a three-parton final state $\gamma^* \to q\bar{q}g$, momentum conservation in the center-of-mass frame gives $\mathbf{p}_1 + \mathbf{p}_2 + \mathbf{p}_3 = 0$, implying that the three momenta are coplanar. We define the energy fractions
\begin{equation}
x_i = \frac{2E_i}{Q} = \frac{2p_i \cdot q}{Q^2}\,,
\end{equation}
where $q^\mu$ is the total four-momentum with $q^2 = Q^2$. Energy conservation gives $x_1 + x_2 + x_3 = 2$, with each $x_i \in (0,1)$ for massless partons.

It is convenient to use the normalized invariant masses
\begin{equation}
s_{ij} = \frac{(p_i + p_j)^2}{Q^2} = \frac{2p_i \cdot p_j}{Q^2}\,,
\end{equation}
which for massless partons satisfy $s_{ij} = 1 - x_k$ where $\{i,j,k\}$ is a permutation of $\{1,2,3\}$. The constraint $x_1 + x_2 + x_3 = 2$ becomes
\begin{equation}
\label{eq:sij-constraint}
s_{12} + s_{13} + s_{23} = 1\,.
\end{equation}

For three massless partons, the C-parameter takes the form
\begin{equation}
\label{eq:C-3body}
C = \frac{6\, s_{12}\, s_{13}\, s_{23}}{(1-s_{12})(1-s_{13})(1-s_{23})} = \frac{6(1-x_1)(1-x_2)(1-x_3)}{x_1 x_2 x_3}\,.
\end{equation}
The second form uses the relations $1-x_i = s_{jk}$ and $x_i = 1 - s_{jk} = s_{ij} + s_{ik}$.

\subsection{The shoulder at \texorpdfstring{$C = 3/4$}{C = 3/4}}

We now determine the maximum value of $C$ for three-parton states. Maximizing Eq.~\eqref{eq:C-3body} subject to the constraint~\eqref{eq:sij-constraint} and positivity $s_{ij} > 0$, we use Lagrange multipliers:
\begin{equation}
\frac{\partial}{\partial s_{12}}\Big[ 6 s_{12} s_{13} s_{23} - \lambda(s_{12} + s_{13} + s_{23} - 1) \Big] = 0\,.
\end{equation}
This gives $6 s_{13} s_{23} = \lambda$, and similarly for cyclic permutations. The solution is
\begin{equation}
\label{eq:symmetric-point}
s_{12} = s_{13} = s_{23} = \frac{1}{3}\,,
\end{equation}
which corresponds to equal energy fractions $x_1 = x_2 = x_3 = 2/3$. At this \emph{symmetric point}, the C-parameter achieves its three-parton maximum:
\begin{equation}
C_{\rm max}^{(3)} = \frac{6 \times (1/3)^3}{(2/3)^3} = \frac{3}{4}\,.
\end{equation}

The symmetric configuration corresponds to three partons with equal energies $E_i = Q/3$ separated by $120^\circ$ angles---a geometry resembling the Mercedes-Benz logo, hence the name ``Mercedes'' commonly used in the literature for this trijet topology. Choosing coordinates with the event plane as the $xz$-plane, the four-momenta are
\begin{align}
\label{eq:mercedes-momenta}
p_1^\mu &= \frac{Q}{3}(1, 0, 0, 1)\,, \nonumber\\
p_2^\mu &= \frac{Q}{3}\Big(1, 0, \frac{\sqrt{3}}{2}, -\frac{1}{2}\Big)\,, \\
p_3^\mu &= \frac{Q}{3}\Big(1, 0, -\frac{\sqrt{3}}{2}, -\frac{1}{2}\Big)\,. \nonumber
\end{align}

\subsection{Leading-order distribution: setup and kinematics}
\label{sec:LO-setup}

At leading order in $\as$, the C-parameter distribution comes from $\gamma^* \to q\bar{q}g$:
\begin{equation}
\label{eq:LO-dist}
\frac{1}{\sigma_0}\frac{d\sigma^{\rm LO}}{dC} = \frac{\as}{2\pi} A(C) \, \Theta\!\left(\frac{3}{4} - C\right)\Theta(C)\,,
\end{equation}
where $\sigma_0$ is the Born cross section for $e^+e^- \to q\bar{q}$, and $A(C)$ is the coefficient function that includes the color factor $\CF$ from the matrix element. We now derive the exact analytical form of $A(C)$, following the approach of Gardi and Magnea~\cite{Gardi:2003iv}.

Consider a three-parton final state with massless quark and antiquark momenta $p_1$, $p_2$, and gluon momentum $p_3$, with total momentum $q$ satisfying $q^2 = Q^2$. We introduce the energy fractions in the center-of-mass frame:
\begin{equation}
x_1 = \frac{2p_1 \cdot q}{Q^2}\,, \qquad x_2 = \frac{2p_2 \cdot q}{Q^2}\,, \qquad x_3 = \frac{2p_3 \cdot q}{Q^2} = 2 - x_1 - x_2\,,
\end{equation}
where energy-momentum conservation requires $x_1 + x_2 + x_3 = 2$ and phase space restricts $0 \le x_i \le 1$. In terms of the energy fractions, the C-parameter for three massless partons is
\begin{equation}
\label{eq:C-x1x2}
C(x_1, x_2) = \frac{6(1-x_1)(1-x_2)(x_1 + x_2 - 1)}{x_1 x_2 (2 - x_1 - x_2)}\,.
\end{equation}

The LO coefficient function is obtained by integrating the squared matrix element
\begin{equation}
\label{eq:ME-squared}
|\mathcal{M}|^2 \propto \CF \frac{x_1^2 + x_2^2}{(1-x_1)(1-x_2)}
\end{equation}
over the three-body phase space with the constraint $C(x_1, x_2) = C$. Implementing the constraint via a delta function, the distribution takes the form
\begin{equation}
\label{eq:F0-integral}
\frac{d\sigma}{dC} \propto \int dx_1\, dx_2 \; \frac{x_1^2 + x_2^2}{(1-x_1)(1-x_2)} \; \delta\big(C - C(x_1, x_2)\big)\,.
\end{equation}
The delta function constrains the integration to a curve in the $(x_1, x_2)$ plane. After eliminating one variable using the constraint and performing a suitable change of variables, the remaining integral over the gluon energy fraction $x_3$ involves square roots of cubic polynomials---a signature of \textbf{elliptic integrals}.

\subsection{Computation of \texorpdfstring{$A(3/4)$}{A(3/4)} and the step discontinuity}
\label{sec:step-discontinuity}

The C-parameter distribution exhibits a step discontinuity at $C = 3/4$: the coefficient function $A(C)$ approaches a finite, nonzero value as $C \to 3/4^-$, unlike thrust where $A(\tau) \to 0$ as $\tau \to 1/3^-$. This difference has important consequences for matching across the Sudakov shoulder, as we will discuss in Section~\ref{sec:resummation}. Here we compute $A(3/4)$ directly by examining the phase space structure near the Mercedes configuration, illuminating the origin of this qualitative difference.

Near Mercedes, we parameterize deviations from the symmetric point using the invariants:
\begin{equation}
s_{12} = \frac{1}{3} + s\,, \qquad s_{13} = \frac{1}{3} + t\,, \qquad s_{23} = \frac{1}{3} - s - t\,,
\end{equation}
where $s_{ij} = (p_i + p_j)^2/Q^2 = 1 - x_k$, and the constraint $s_{12} + s_{13} + s_{23} = 1$ is automatically satisfied. The Mercedes configuration corresponds to $s = t = 0$. The matrix element at this point evaluates to
\begin{equation}
\frac{x_1^2 + x_2^2}{(1-x_1)(1-x_2)}\Big|_{\rm Merc} = \frac{2(2/3)^2}{(1/3)^2} = 8\,.
\end{equation}

\paragraph{Thrust: linear observable.}
For thrust, near Mercedes in the region where $s_{12}$ is minimal, the observable is simply $\tau = s_{12} = 1/3 + s$. The coefficient function is
\begin{equation}
A_\tau(\tau) = 8\CF \int ds\, dt\, \delta\!\left(\tau - \frac{1}{3} - s\right) \Theta(t - s)\, \Theta(-s - t)\,,
\end{equation}
where the step functions enforce $s_{12} \leq s_{13}$ and $s_{12} \leq s_{23}$. After integrating over $s$ using the delta function, the constraint restricts $t \in [s, -2s]$ with $s = \tau - 1/3 < 0$:
\begin{equation}
A_\tau(\tau) = 8\CF \int_{\tau - 1/3}^{-2(\tau - 1/3)} dt = 8\CF \cdot 3\left(\frac{1}{3} - \tau\right) \xrightarrow{\tau \to 1/3} 0\,.
\end{equation}
The distribution vanishes linearly at the shoulder because thrust is a \emph{linear} function of the phase space coordinates near Mercedes.

\paragraph{C-parameter: quadratic observable.}
For the C-parameter, expanding around the Mercedes configuration gives
\begin{equation}
\label{eq:C-Mercedes-expansion}
C = \frac{3}{4} - \frac{81}{16}(s^2 + st + t^2) + \order{\epsilon^3}\,.
\end{equation}
Crucially, the Mercedes point is a \emph{critical point} of the C-parameter: $\nabla C = 0$ at $s = t = 0$. The observable is quadratic in deviations from Mercedes, not linear.

The coefficient function becomes
\begin{equation}
A_C(C) = 8\CF \int ds\, dt\, \delta\!\left(\frac{3}{4} - C - \frac{81}{16}(s^2 + st + t^2)\right).
\end{equation}
Defining the quadratic form $\mathcal{Q}(s,t) = s^2 + st + t^2$, the constraint is $\mathcal{Q} = (16/81)(3/4 - C)$.

To evaluate the integral, we diagonalize the quadratic form. Writing $Q = (s + t/2)^2 + 3t^2/4$ suggests the substitution $u = s + t/2$, $v = \sqrt{3}t/2$, giving $Q = u^2 + v^2 \equiv r^2$ with Jacobian
\begin{equation}
ds\, dt = \frac{2}{\sqrt{3}}\, du\, dv = \frac{2}{\sqrt{3}}\, r\, dr\, d\theta\,.
\end{equation}
The integral becomes
\begin{equation}
A_C(C) = 8\CF \cdot \frac{2}{\sqrt{3}} \int_0^{2\pi} d\theta \int_0^\infty r\, dr\, \delta\!\left(\tfrac{3}{4} - C - \frac{81}{16}r^2\right).
\end{equation}
Using the identity $\delta(a - \alpha r^2) = \frac{1}{2\alpha r_0}\delta(r - r_0)$ where $r_0 = \sqrt{a/\alpha}$, the factors of $r$ cancel exactly:
\begin{equation}
A_C(C) = 8\CF \cdot \frac{2}{\sqrt{3}} \cdot 2\pi \cdot \frac{1}{2 \cdot 81/16} = \frac{256\sqrt{3}\pi}{243}\CF\,.
\end{equation}
This gives the result:
\begin{equation}
\label{eq:A34-direct}
A\!\left(\frac{3}{4}\right) = \frac{256\sqrt{3}\pi}{243}\CF \approx 7.64
\end{equation}
for $\CF = 4/3$.

\paragraph{Physical interpretation.}
The key difference between thrust and C-parameter lies in the structure of the observable near Mercedes:
\begin{itemize}
\item \textbf{Thrust:} $\tau - 1/3 \propto r$ (linear in phase space distance). The phase space volume at fixed $\tau$ shrinks like the length of a line segment, proportional to $r \to 0$.
\item \textbf{C-parameter:} $3/4 - C \propto r^2$ (quadratic in phase space distance). The phase space volume at fixed $C$ shrinks like the circumference of a circle, proportional to $r$, but the Jacobian $\partial C/\partial r \propto r$ also vanishes, and the two effects cancel exactly.
\end{itemize}
This cancellation is analogous to the density of states for a free particle in two dimensions. For a 2D system with energy $E \propto v^2$, the density of states $g(E)$ is constant even as $E \to 0$: the probability of finding a particle with kinetic energy less than $\epsilon$ scales linearly with $\epsilon$, not quadratically. The Mercedes configuration is a critical point of $C(\{s_{ij}\})$, making the C-parameter locally equivalent to ``energy'' in this analogy, while thrust is analogous to ``speed'' $v$ whose distribution vanishes at $v = 0$.

\subsection{Exact analytical formula for \texorpdfstring{$A(C)$}{A(C)}}
\label{sec:LO-exact}

The integration in Eq.~\eqref{eq:F0-integral} can be performed analytically. The result, derived in Ref.~\cite{Gardi:2003iv}, expresses the LO coefficient as a linear combination of complete elliptic integrals. Following that reference, we define
\begin{equation}
\label{eq:cGM-def}
\cGM \equiv \frac{C}{6}\,,
\end{equation}
so that the physical support at LO is $0 < \cGM < 1/8$. The coefficient function is
\begin{equation}
\label{eq:A-from-F0}
A(C) = \frac{1}{6}\,F_0(\cGM)\,, \qquad 0 < C < \frac{3}{4}\,,
\end{equation}
where the ``characteristic function'' $F_0(\cGM)$ is given by
\begin{equation}
\label{eq:F0-elliptic}
F_0(\cGM) = f_0(\cGM)\,\mathbf{K}\big(m_0(\cGM)\big) + e_0(\cGM)\,\mathbf{E}\big(m_0(\cGM)\big) + p_0(\cGM)\,\boldsymbol{\Pi}\big(n_0(\cGM), m_0(\cGM)\big)\,.
\end{equation}
Here $\mathbf{K}$, $\mathbf{E}$, and $\boldsymbol{\Pi}$ are the complete elliptic integrals of the first, second, and third kind, respectively, defined by
\begin{align}
\mathbf{K}(m) &= \int_0^{\pi/2} \frac{d\phi}{\sqrt{1 - m\sin^2\phi}}\,, \label{eq:ellipK} \\
\mathbf{E}(m) &= \int_0^{\pi/2} d\phi\,\sqrt{1 - m\sin^2\phi}\,, \label{eq:ellipE} \\
\boldsymbol{\Pi}(n, m) &= \int_0^{\pi/2} \frac{d\phi}{(1 - n\sin^2\phi)\sqrt{1 - m\sin^2\phi}}\,. \label{eq:ellipPi}
\end{align}

The modulus, characteristic parameter, and coefficient functions are~\cite{Gardi:2003iv}:\footnote{Eq.~(A.17) of Ref.~\cite{Gardi:2003iv} contains a typographical error in the expression for $e_0(\cGM)$: the denominator is printed as $\sqrt{2\cGM}\,(1+\cGM)^3$ but should be $2\cGM\,(1+\cGM)^3$. Equivalently, the published formula is too small by a factor of $\sqrt{\cGM}$. We give the corrected expression here.}
\begin{align}
m_0(\cGM) &= \frac{2\sqrt{1-8\cGM}}{1 - 4\cGM(1+2\cGM) + \sqrt{1-8\cGM}}\,, \label{eq:m0} \\
n_0(\cGM) &= \frac{4\sqrt{1-8\cGM}}{\big(1 + \sqrt{1-8\cGM}\big)^2}\,, \label{eq:n0} \\
f_0(\cGM) &= \frac{4\sqrt{2}\,\big(1 - 2\cGM(2+\cGM)\big)}{(1+\cGM)^3\,\sqrt{1 - 4\cGM(1+2\cGM) + \sqrt{1-8\cGM}}}\,, \label{eq:f0} \\
e_0(\cGM) &= -\frac{3(1+2\cGM)\sqrt{2}\,\sqrt{1 - 4\cGM(1+2\cGM) + \sqrt{1-8\cGM}}}{2\cGM\,(1+\cGM)^3}\,, \label{eq:e0} \\
p_0(\cGM) &= \frac{\sqrt{2}\,(2 + \cGM + 2\cGM^2)\,\big(1 - \sqrt{1-8\cGM}\big)^2}{\cGM(1+\cGM)^3\,\sqrt{1 - 4\cGM(1+2\cGM) + \sqrt{1-8\cGM}}}\,. \label{eq:p0}
\end{align}
These expressions are valid for $0 < \cGM < 1/8$, i.e., $0 < C < 3/4$.

\subsection{Asymptotic behavior and numerical validation}
\label{sec:LO-asymptotics}

\begin{figure}[t]
\centering
\includegraphics[width=\textwidth]{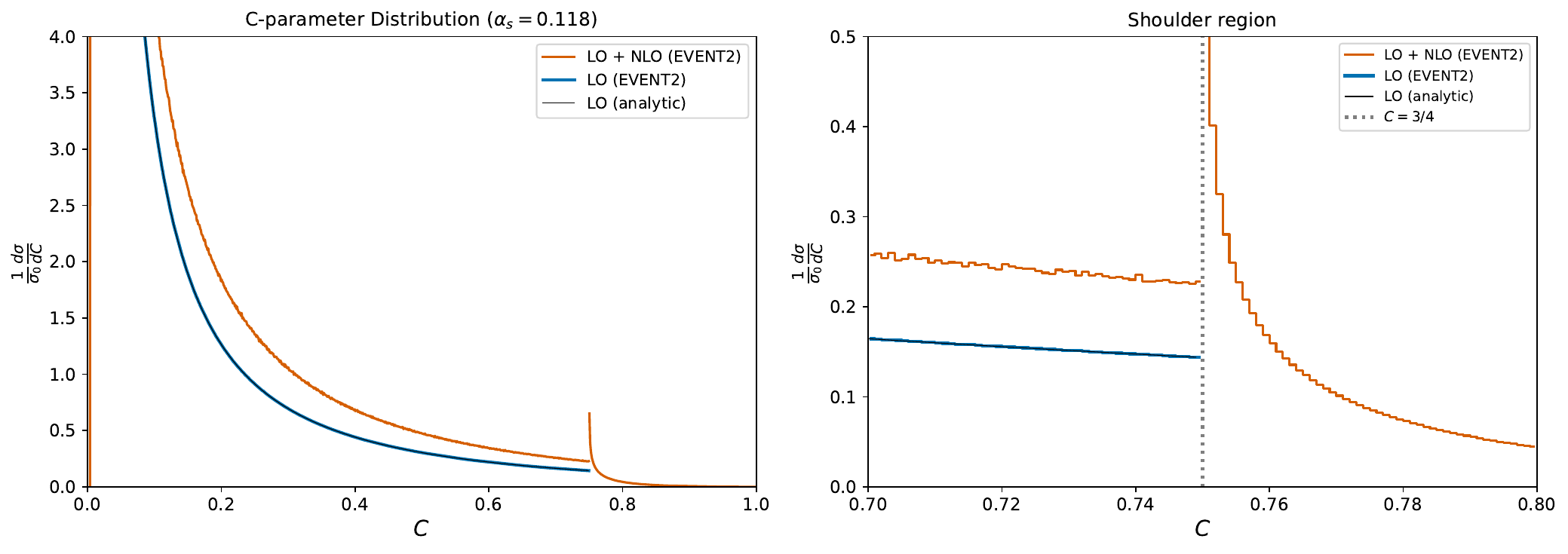}
\caption{C-parameter distribution at LO and NLO from EVENT2 Monte Carlo (with $\as = 0.118$). Left: full distribution showing the characteristic $1/C$ divergence at small $C$ and the step discontinuity at the shoulder $C = 3/4$. The LO distribution $(\as/2\pi)A(C)$ from Eq.~\eqref{eq:LO-dist} agrees precisely with the exact analytical formula (black curve) from Eqs.~\eqref{eq:A-from-F0}--\eqref{eq:p0}. Right: zoom on the shoulder region, showing the NLO spike just above $C = 3/4$ from unresummed Sudakov logarithms, while the LO distribution vanishes exactly for $C > 3/4$.}
\label{fig:LO-comparison}
\end{figure}

The exact formula Eq.~\eqref{eq:F0-elliptic} can be expanded in various limits. As $\cGM \to 0$ (soft/collinear region), the elliptic integrals simplify, yielding~\cite{Gardi:2003iv}
\begin{equation}
\label{eq:F0-smallc}
F_0(\cGM) = -\frac{3 + 4\ln \cGM}{\cGM} + 1 - 28\ln \cGM + \order{\cGM\ln \cGM}\,.
\end{equation}
This reproduces the known singular structure from soft and collinear emissions. As $\cGM \to 1/8$ (approaching the shoulder), the elliptic modulus $m_0 \to 0$ and the elliptic integrals approach $\mathbf{K}(0) = \mathbf{E}(0) = \pi/2$. Evaluating the limit confirms Eq.~\eqref{eq:A34-direct}:
\begin{equation}
\label{eq:A34}
\boxed{A\!\left(\frac{3}{4}\right) = \frac{256\pi\sqrt{3}}{243}\, \CF}
\end{equation}
Figure~\ref{fig:LO-comparison} compares the exact analytical formula with EVENT2 Monte Carlo results, demonstrating excellent agreement across the entire kinematic range $0 < C < 3/4$. The figure also shows the NLO distribution, which extends beyond the shoulder and exhibits the Sudakov spike at $C = 3/4^+$.

\section{SCET Factorization Theorem}
\label{sec:factorization}

We now derive the factorization theorem for the C-parameter distribution near the shoulder using soft-collinear effective theory. The key insight is that configurations just above $C = 3/4$ are perturbations of the symmetric Mercedes trijet, with the excess coming from soft and collinear radiation.

\subsection{Soft and collinear radiation near the shoulder}

The symmetric trijet configuration saturates the bound $C \le 3/4$ for three massless partons. To access the region $C > 3/4$, additional radiation must carry the system away from this symmetric point.

The SCET description involves three types of modes:
\begin{itemize}
\item \textbf{Hard modes} carry momenta of order $Q$ and are integrated out when matching QCD onto SCET, producing the hard function $H(\{s_{ij}\},\mu)$ that encodes the production of the underlying trijet. The hard function depends on the dipole invariants $s_{ij} = (p_i + p_j)^2$; at the Mercedes configuration, $s_{ij} = Q^2/3$ for all pairs.
\item \textbf{Collinear modes} are boosted along one of the three jet directions and build up the jet invariant masses. The collinear radiation from each jet is described by a jet function $J_i^C(\ell,\mu)$.
\item \textbf{Soft modes} have all momentum components of order $Qc$ and contribute to $c$ through out-of-plane momentum. Their contribution is encoded in the soft function $S(k,\mu)$, where $k$ is the dimensionful soft measurement variable.
\end{itemize}

The three jet directions are specified by the symmetric trijet configuration:
\begin{align}
n_1^\mu &= (1, 0, 0, 1)\,, \nonumber\\
n_2^\mu &= (1, 0, \tfrac{\sqrt{3}}{2}, -\tfrac{1}{2})\,, \\
n_3^\mu &= (1, 0, -\tfrac{\sqrt{3}}{2}, -\tfrac{1}{2})\,, \nonumber
\end{align}
corresponding to quark, antiquark, and gluon jets separated by $120^\circ$ angles.

\subsection{Measurement decomposition}

A crucial feature enabling factorization is that the observable decomposes additively at leading power. We derive the soft and collinear contributions starting from the pair-sum formula Eq.~\eqref{eq:C-def}. For massless particles, $|\mathbf{p}_i| = E_i$ and $\sum_k |\mathbf{p}_k| = Q$, so the C-parameter becomes
\begin{equation}
\label{eq:C-angle-formula}
C = \frac{3}{Q^2}\sum_{i<j} E_i E_j \sin^2\theta_{ij}\,.
\end{equation}

\textbf{Soft contribution.} Consider adding a soft gluon with energy $\omega \ll Q$ to the Mercedes configuration. Energy conservation requires the hard partons to share energy $(Q-\omega)$, giving each hard parton energy $(Q-\omega)/3$. Using Eq.~\eqref{eq:C-angle-formula}, the C-parameter becomes
\begin{equation}
C = \frac{3}{Q^2}\left[3 \times \left(\frac{Q-\omega}{3}\right)^2 \sin^2(120^\circ) + \sum_{i=1}^{3} E_i \omega \sin^2\theta_{i,\text{soft}}\right],
\end{equation}
where $\theta_{i,\text{soft}}$ is the angle between the soft gluon and hard parton $i$. The first term is the contribution from hard parton pairs, and the second from hard-soft pairs.

For the Mercedes geometry, we parameterize the soft gluon direction by its polar angle $\psi$ from the event plane. The sum over hard-soft angles satisfies $\sum_i\sin^2\theta_{i,\text{soft}} = \frac{3}{2}(1+\sin^2\psi)$, where the $3/2$ comes from the symmetric trijet geometry. Expanding to linear order in $\omega/Q$:
\begin{equation}
C - \frac{3}{4} = \frac{\omega}{Q}\left[\frac{3}{2}(1+\sin^2\psi) - \frac{3}{2}\right] = \frac{3\omega}{2Q}\sin^2\psi\,.
\end{equation}
Defining the shoulder variable
\begin{equation}
\label{eq:c-def}
c \equiv \frac{8}{3}\left(C - \frac{3}{4}\right),
\end{equation}
the soft contribution from a single gluon is
\begin{equation}
\label{eq:delta-soft}
c_{\rm soft} = \frac{4\omega\sin^2\psi}{Q} = \frac{4k_\perp^2}{\omega Q}\,,
\end{equation}
where $k_\perp = \omega\sin\psi$ is the momentum component perpendicular to the event plane. In-plane emission ($\psi = 0$) gives $c_{\rm soft} = 0$: soft gluons emitted within the event plane do not contribute to $C > 3/4$. The soft measurement is additive across multiple emissions:
\begin{equation}
c_{\rm soft}^{\rm total} = \frac{4}{Q}\sum_{\text{soft }k} \frac{k_\perp^2}{k^0}\,.
\end{equation}

\textbf{Collinear contribution.} Consider a collinear splitting of jet $j$ (with energy $E_j = Q/3$) into two partons with energies $zE_j$ and $(1-z)E_j$ and opening angle $\theta \ll 1$. Let $\phi$ denote the azimuthal angle of the splitting plane about the jet axis, measured relative to the event plane. The two daughter partons have directions
\begin{equation}
\hat{p}_1 = \hat{n}_j + (1-z)\theta(\cos\phi\,\hat{e}_\perp + \sin\phi\,\hat{e}_\parallel)\,, \quad
\hat{p}_2 = \hat{n}_j - z\theta(\cos\phi\,\hat{e}_\perp + \sin\phi\,\hat{e}_\parallel)\,,
\end{equation}
where $\hat{e}_\perp$ is perpendicular to the event plane and $\hat{e}_\parallel$ is in the event plane but perpendicular to $\hat{n}_j$.

Using Eq.~\eqref{eq:C-angle-formula}, the C-parameter receives contributions from three types of pairs: (i) the collinear pair itself, (ii) each collinear parton with hard partons from other jets in the event plane, and (iii) the hard partons among themselves. For (i), $\sin^2\theta_{12} = \theta^2$, giving $\frac{3}{Q^2}z(1-z)E_j^2\theta^2 = \frac{m^2}{3Q^2}$ where $m^2 = z(1-z)E_j^2\theta^2$ is the jet invariant mass. For (iii), the hard-hard contribution is unchanged at $3/4$ to leading order in $\theta$.

The key contribution is (ii). Consider the angle between collinear parton 1 and a hard parton $k$ in a different jet. The Mercedes jets are at $120^\circ$ from jet $j$, so $\hat{n}_j \cdot \hat{n}_k = -1/2$. The angle shift is
\begin{equation}
\sin^2\theta_{1k} - \sin^2(120^\circ) = -2\cos(120^\circ)\,\hat{n}_k \cdot \delta\hat{p}_1 = (1-z)\theta\sin\phi\,(\hat{n}_k \cdot \hat{e}_\parallel)\,,
\end{equation}
where only the in-plane component $\sin\phi$ contributes since $\hat{n}_k \cdot \hat{e}_\perp = 0$ (the hard partons lie in the event plane). Summing over all hard-collinear pairs and using $(\hat{n}_k \cdot \hat{e}_\parallel)^2 = 3/4$ from the Mercedes geometry:
\begin{equation}
\delta C = \frac{3}{Q^2} \times 2 \times \frac{Q}{3} \times \frac{Q}{3} \times \frac{9}{2} \times z(1-z)\theta^2\sin^2\phi = \frac{9}{2}\sin^2\phi\,\frac{m^2}{Q^2}\,.
\end{equation}
Converting to the shoulder variable $c = \frac{8}{3}(C - 3/4)$:
\begin{equation}
\label{eq:delta-coll}
c_{\rm coll} = 12\sin^2\phi \cdot \frac{m^2}{Q^2}\,.
\end{equation}
In-plane splittings ($\phi = 0$) give $c_{\rm coll} = 0$, while maximally out-of-plane splittings ($\phi = \pi/2$) give $c_{\rm coll} = 12m^2/Q^2$. For multiple collinear emissions within a jet, each emission $a$ with virtuality $p_a^2$ and azimuthal angle $\phi_a$ contributes independently:
\begin{equation}
c_{\rm coll}^{\rm jet} = \frac{12}{Q^2}\sum_a p_a^2 \sin^2\phi_a\,.
\end{equation}

\textbf{Additivity and mode separation.} Soft and collinear contributions separate because: (i) both enter linearly in the perturbation at leading power, and (ii) soft-collinear modes (soft gluons collinear to a jet) have $k_\perp \to 0$ since all jet directions lie in the event plane. The total shoulder variable decomposes as
\begin{equation}
\label{eq:additive}
c = c_{\rm coll} + c_{\rm soft}\,.
\end{equation}

The characteristic momentum scalings in light-cone coordinates $(p^+, p^-, p_\perp)$ relative to a jet direction are:
\begin{align}
\text{Collinear:} \quad & p_c \sim Q(c, 1, \sqrt{c})\,, \quad p_c^2 \sim Q^2 c\,, \nonumber\\
\text{Soft:} \quad & p_s \sim Q(c, c, c)\,,\quad\;\; p_s^2 \sim Q^2 c^2\,.
\end{align}
These scalings determine the canonical scale hierarchy:
\begin{equation}
\label{eq:scale-hierarchy}
\mu_S = Qc \ll \mu_J = Q\sqrt{c} \ll \mu_H = Q\,.
\end{equation}
Large logarithms $\ln c$ arise from ratios of these scales; resummation sums them by evolving each function from its canonical scale to a common scale.

\subsection{The factorization theorem}
\label{sec:factorization-theorem}

Combining mode separation with measurement decomposition, the cross section factorizes. The SCET kernel for the C-parameter shoulder is:
\begin{multline}
\label{eq:kernel-convolution}
K(c,Q/\mu) = H(Q,\mu) \int\! dm_1^2\, dm_2^2\, dm_3^2\, dk\; J^C_q(m_1^2,\mu)\, J^C_{\bar{q}}(m_2^2,\mu)\, J^C_g(m_3^2,\mu)\, \\
\times\, S(k,\mu)\;\delta\!\left(cQ^2 - m_1^2 - m_2^2 - m_3^2 - Qk\right).
\end{multline}
We now define the perturbative ingredients.

\paragraph{Hard function.}
The hard function $H(Q,\mu)$ is the squared Wilson coefficient from matching QCD onto the three-jet SCET operator at Mercedes kinematics ($s_{ij} = Q^2/3$). It encodes the short-distance production of the underlying trijet configuration and is identical to the hard function for thrust and heavy jet mass at the symmetric point.

\paragraph{C-shoulder jet functions.}
The C-shoulder jet functions $J^C_q(m^2,\mu)$ and $J^C_g(m^2,\mu)$ are defined as vacuum matrix elements of collinear fields with the C-parameter measurement operator. For a quark jet:
\begin{equation}
\label{eq:JC-def-fact}
J^C_q(m^2,\mu) = \frac{1}{2N_c}\,{\rm tr}\,\langle 0 | \bar{\chi}_n\, \delta(m^2 - \widehat{M}^C_J)\,\frac{\not{\!\bar{n}}}{2}\, \chi_n | 0 \rangle\,,
\end{equation}
where $\chi_n$ is the gauge-invariant collinear quark field in SCET, and the C-parameter measurement operator is
\begin{equation}
\label{eq:MC-def-fact}
\widehat{M}^C_J = 12\sum_{a} \frac{(p_{a\perp}^x)^2}{\bar{n}\cdot p_a}\,.
\end{equation}
Here $p_{a\perp}^x$ is the component of the transverse momentum perpendicular to the event plane, and the sum runs over all collinear particles in the jet. The factor of 12 is the geometric factor from the Mercedes configuration derived in Eq.~\eqref{eq:delta-coll}. The variable $m^2$ has mass dimension two and represents the C-parameter weighted jet mass. For a gluon jet:
\begin{equation}
J^C_g(m^2,\mu) = \frac{\omega}{2(N_c^2-1)}\,\langle 0 | \mathcal{B}_{n\perp}^\mu\, \delta(m^2 - \widehat{M}^C_J)\, \mathcal{B}_{n\perp\mu} | 0 \rangle\,,
\end{equation}
where $\mathcal{B}_{n\perp}^\mu$ is the gauge-invariant collinear gluon field. The C-shoulder jet functions differ from the standard inclusive jet functions by incorporating the azimuthal weighting $\sin^2\phi$ intrinsic to the C-parameter measurement.

\paragraph{Soft function.}
The soft function $S(k,\mu)$ is the vacuum matrix element of soft Wilson lines along the three jet directions with the C-parameter measurement operator:
\begin{equation}
\label{eq:soft-def}
S(k,\mu) = \frac{1}{N_c}\,{\rm Tr}\,\langle 0| \bar{T}\{S_{n_3}^\dagger S_{n_2}^\dagger S_{n_1}^\dagger\}\,\delta(k - \widehat{M}_S)\,T\{S_{n_1} S_{n_2} S_{n_3}\}|0\rangle\,,
\end{equation}
where $S_{n_i}$ are soft Wilson lines in the appropriate color representations (fundamental for quarks, adjoint for the gluon) and
\begin{equation}
\widehat{M}_S = 4\sum_{\text{soft }k} \frac{k_\perp^2}{k^0}
\end{equation}
is the soft measurement operator derived in Eq.~\eqref{eq:delta-soft}. The variable $k$ has mass dimension one and scales as $k \sim Qc$. At tree level, $K^{(0)}(c) = \delta(c)$ since all functions are normalized to delta functions.

\subsection{The resummed shoulder cross section}
\label{sec:resummed-xsec}

The full factorization must include the integral over hard phase space. Near Mercedes, we parameterize deviations by $(s,t)$ as in Section~\ref{sec:step-discontinuity}, giving $C_{\rm hard} = 3/4 - (81/16) \mathcal{Q}(s,t)$ where $\mathcal{Q}(s,t) = s^2 + st + t^2$. In terms of the shoulder variable Eq.~\eqref{eq:c-def}, this corresponds to $c_{\rm hard} = -(27/2) \mathcal{Q}(s,t)$.

Let $\mathcal{H}_{\rm Born}(s,t)$ denote the Born-level phase-space measure (the squared matrix element without the hard Wilson coefficient). The factorized cross section is:
\begin{equation}
\label{eq:factorization-full}
\frac{d\sigma}{dc} = \int\! ds\,dt\; \mathcal{H}_{\rm Born}(s,t)\, \int\! dc'\; K(c',\mu)\; \delta\big(c + \tfrac{27}{2} \mathcal{Q}(s,t) - c'\big)\,,
\end{equation}
where $K(c',\mu)$ is the full SCET kernel from Eq.~\eqref{eq:kernel-convolution}.

The crucial observation is that the hard phase space integral with the quadratic constraint produces a \emph{constant} for any $c' > c$:
\begin{equation}
\label{eq:jacobian-identity}
\int\! ds\,dt\; \mathcal{H}_{\rm Born}(s,t)\; \delta\big(c + \tfrac{27}{2} \mathcal{Q}(s,t) - c'\big) = A_{\rm Born}(3/4)\, \Theta(c' - c)\,,
\end{equation}
where $A_{\rm Born}(3/4) = (256\pi\sqrt{3}/243)\CF$ is the Born coefficient from Eq.~\eqref{eq:A34-direct}. This is the same Jacobian cancellation that produced the finite LO coefficient in Section~\ref{sec:step-discontinuity}. Substituting into the factorized cross section:
\begin{equation}
\frac{d\sigma}{dc} = A_{\rm Born}(3/4) \int_c^\infty\! dc'\; K(c',\mu)\,.
\end{equation}

We define the cumulant as the integral of the kernel from $0$ to $c$:
\begin{equation}
\label{eq:cumulant-R}
R(c,\mu) \equiv \int_0^c dc'\; K(c',\mu)\,.
\end{equation}
At tree level, $K^{(0)}(c') = \delta(c')$, so $R^{(0)}(c) = \theta(c)$. Beyond tree level, radiative corrections smooth out this step function into a continuous Sudakov form factor. The key property is that after resummation, Sudakov suppression gives $R(c) \to 0$ as $c \to 0^+$---the logarithmically-enhanced contributions are exponentiated into a vanishing factor at the shoulder.

Naively, one might write the cross section as
\begin{equation}
\label{eq:naive-formula}
\frac{d\sigma}{dc} \propto A_{\rm Born}(3/4) \int_c^\infty dc'\, K(c') = A_{\rm Born}(3/4)\left[\int_0^\infty dc'\, K(c') - R(c)\right].
\end{equation}
However, the integral $\int_0^\infty K(c')\,dc'$ is formally divergent---the kernel $K(c')$ contains plus distributions and is not a normalized probability density. This divergence is a $c$-independent constant.

Rather than tracking this divergent constant through the matching, we instead organize the cross section using the factor $(1 - R(c))$:
\begin{equation}
\label{eq:master-formula}
\frac{1}{\sigma_0}\frac{d\sigma}{dc}\bigg|_{c>0} = \frac{\as}{2\pi}\,A_{\rm Born}(3/4)\, \big(1 - R(c)\big) + \sigma_{\rm NS}(c)\,,
\end{equation}
where $\sigma_{\rm NS}$ is the non-singular remainder from matching to fixed order. This formulation has two key advantages. First, without matching ($\sigma_{\rm NS} = 0$), the resummation correctly reduces to the LO limit at the shoulder: as $c \to 0^+$, Sudakov suppression gives $R(c) \to 0$, so the distribution approaches $(\as/2\pi)\,A_{\rm Born}(3/4)$, precisely the LO value from below. Second, with matching, the full expression is formally $\order{\as^2}$ or smaller in the $C > 3/4$ region: at tree level $R(c) = 1$ for $c > 0$, so the $(1-R)$ piece vanishes above the shoulder and the fixed-order expansion starts at $\order{\as^2}$, as expected from the NLO structure. The divergent integral $\int_0^\infty K(c')\,dc'$ is absorbed into the definition of $\sigma_{\rm NS}$ through the matching procedure and never appears explicitly.

The details of matching to fixed order, including the treatment of the non-singular contribution $\sigma_{\rm NS}$, are discussed in Section~\ref{sec:matching-theory}.

\subsection{Comparison with heavy jet mass}

The factorization has the same general structure as heavy jet mass (HJM)~\cite{Bhattacharya:2022vrh}: $d\sigma/dc = H \otimes J_q^C \otimes J_{\bar{q}}^C \otimes J_g^C \otimes S$. However, both the jet and soft functions differ from those in thrust or HJM. The hard function $H(\{s_{ij}\},\mu)$ is shared, depending on the dipole invariants $s_{ij}$ which at Mercedes take the symmetric value $s_{ij} = Q^2/3$.

The C-shoulder jet functions $J_i^C$ differ from the inclusive jet functions appearing in thrust/HJM. As discussed in Section~\ref{sec:intro}, the inclusive jet function measures only the total invariant mass of collinear radiation, while the C-shoulder jet function measures $\sum_a (p_{a\perp}^x)^2/E_a$---the azimuthally-weighted transverse momentum squared divided by energy. This measurement projects onto a single transverse momentum component (perpendicular to the event plane), introducing the $\sin^2\phi$ dependence characteristic of the C-parameter. The one-loop C-shoulder jet functions can be computed from the inclusive ones by averaging over the azimuthal angle (see Section~\ref{sec:jet-functions} and Appendix~\ref{app:Cjet}).

The soft function also differs fundamentally from thrust/HJM:
\begin{enumerate}
\item \textbf{Universal angular function:} The measurement function is the \emph{same} everywhere---there is no sextant decomposition with different projection vectors.
\item \textbf{Quadratic angular dependence:} The soft contribution is \emph{quadratic} in the out-of-plane momentum $k_\perp$, not a linear projection.
\item \textbf{Vanishing for in-plane radiation:} Soft gluons emitted in the event plane ($k_\perp = 0$) contribute \emph{zero} to $\delta C$ at leading power.
\item \textbf{Homogeneity:} Despite the quadratic angular dependence, the measurement is homogeneous of degree 1 in the soft momentum: $M_S(\lambda k) = \lambda M_S(k)$.
\end{enumerate}

A key simplification for the C-parameter is that all channels give identical contributions. For HJM, ``gluon'' and ``quark'' channels based on which parton is isolated contribute differently because the hemisphere boundary determines the observable. The C-parameter, being a symmetric global sum over all particles, cannot distinguish which parton is isolated---the channel decomposition has no physical significance. This equivalence is exact to all orders by permutation symmetry.

\subsection{Absence of non-global logarithms}

An important feature of the C-parameter is that it is a \textbf{global observable}: every final-state particle contributes to $C$ based on its momentum, with no vetoes or restricted phase-space regions. This ensures that non-global logarithms (NGLs)~\cite{Dasgupta:2001sh} are absent at leading power.

NGLs arise when an observable restricts radiation to a limited phase-space region, allowing soft gluons outside this region to radiate back into it. Classic examples include the ``gap between jets'' and cone-based observables. For such observables, correlated multi-gluon emission generates logarithms not captured by independent-emission resummation.

The C-parameter avoids this problem because:
\begin{enumerate}
    \item It sums over all final-state particles---no particles are excluded.
    \item There are no geometric boundaries (hemispheres, cones) that define ``in'' versus ``out'' regions.
    \item All radiation contributes positively to $C$; there are no cancellations between regions.
\end{enumerate}
This global nature ensures that the factorization theorem Eq.~\eqref{eq:factorization-full} captures all leading-power logarithms, with no additional NGL contributions requiring separate treatment.


\section{Perturbative Ingredients}
\label{sec:ingredients}

We now specify the perturbative ingredients entering the factorization theorem. The hard function is identical to that appearing in the BSZ analysis of thrust and heavy jet mass, but the jet and soft functions are new objects specific to the C-parameter measurement.

\subsection{Hard function}
\label{sec:hard-function}

The hard function $H(Q,\mu)$ is obtained by matching the QCD electromagnetic current onto the three-jet SCET operator. This matching is performed by computing the virtual corrections to the $e^+e^- \to q\bar{q}g$ amplitude in full QCD and in SCET, and taking their ratio. Since SCET reproduces all infrared divergences of QCD, the hard function is infrared-finite and encodes the short-distance physics at scale $Q$.

The threshold endpoint configuration for the C-parameter shoulder is identical to that for thrust and heavy jet mass at the trijet threshold: all three partons have equal energies $E_i = Q/3$ and are separated by $120^\circ$ angles. Since the hard function depends only on the underlying hard kinematics (the dipole invariants $s_{ij} = Q^2/3$ at Mercedes), it is the same as that computed in the BSZ analysis~\cite{Becher:2008cf}. The one-loop hard function can be extracted from the virtual corrections to the trijet amplitude, which are known analytically~\cite{Ellis:1980wv,Giele:1991vf}.

\paragraph{Anomalous dimension.}
The hard function satisfies the RG equation $\mu\,d\ln H/d\mu = \gamma_H$, with anomalous dimension determined by the infrared structure of QCD amplitudes. For a multi-parton amplitude with massless colored partons, the anomalous dimension takes the general dipole form~\cite{Becher:2009th,Gardi:2009qi}:
\begin{equation}
\label{eq:gamma-H-general}
\gamma_H = -\Gcusp(\as)\sum_{i<j} \bm{T}_i \cdot \bm{T}_j \ln\frac{s_{ij}}{\mu^2} + \sum_i \gamma_i(\as)\,,
\end{equation}
where $\bm{T}_i$ are the color generators for parton $i$, $s_{ij} = 2p_i \cdot p_j$ are the dipole invariants, and $\gamma_i$ are the single-parton anomalous dimensions. For a trijet with a quark, antiquark, and gluon, color conservation gives $\bm{T}_q \cdot \bm{T}_{\bar{q}} = -\CF + \CA/2$, $\bm{T}_q \cdot \bm{T}_g = \bm{T}_{\bar{q}} \cdot \bm{T}_g = -\CA/2$, so the total cusp color factor is
\begin{equation}
\sum_{i<j} \bm{T}_i \cdot \bm{T}_j = -\CF + \frac{\CA}{2} - \frac{\CA}{2} - \frac{\CA}{2} = -\CF - \frac{\CA}{2} = -\frac{\mathcal{C}}{2}\,,
\end{equation}
where $\mathcal{C} = 2\CF + \CA$.

At Mercedes, all dipole invariants are equal ($s_{ij} = Q^2/3$), so Eq.~\eqref{eq:gamma-H-general} simplifies. At leading order, the anomalous dimension is
\begin{equation}
\label{eq:gamma-H}
\gamma_H^{(0)} = -6\CF - \beta_0 - 2\mathcal{C}\ln 3\,,
\end{equation}
where the $\ln 3$ arises from the Mercedes geometry: $\ln(s_{ij}/Q^2) = -\ln 3$.

The tree-level hard function is normalized to unity, $H^{(0)} = 1$, so that the Born coefficient $A_{\rm Born}(3/4)$ from Eq.~\eqref{eq:A34-direct} enters as the overall prefactor in the cross section.
The one-loop hard function is
\begin{equation}
H(Q,\mu) = 1 + \frac{\as}{4\pi}\left[-\mathcal{C}\frac{\Gamma_0}{4}L_H^2 - \gamma_H^{(0)} L_H + c_H^1\right] + \order{\as^2}\,,
\end{equation}
where $L_H = \ln(Q^2/\mu^2)$. The one-loop constant for the trijet configuration at the Mercedes point is~\cite{BMSSZ:2023}
\begin{multline}
\label{eq:cH1}
c_H^1 = \CF\left[-\frac{65}{4} + \frac{3\pi^2}{2} - \frac{21}{8}\ln 3 - 10\ln 2\ln 3 + 3\ln^2 3 + 10\,{\rm Li}_2\!\left(\frac{1}{3}\right)\right] \\
+ \CA\left[\frac{3}{4} + \frac{5\pi^2}{4} + \frac{3}{8}\ln 3 + \ln 2\ln 3 - \frac{3}{2}\ln^2 3 - {\rm Li}_2\!\left(\frac{1}{3}\right)\right].
\end{multline}

\subsection{C-shoulder jet function}
\label{sec:jet-functions}

The collinear contribution to the C-parameter at the shoulder involves an azimuthal dependence that is not captured by the standard inclusive jet function. As derived in Section~\ref{sec:factorization}, the collinear contribution from jet $i$ is $c_i = 12\sin^2\phi \cdot m^2/Q^2$, where $m^2$ is the jet invariant mass and $\phi$ is the azimuthal angle of the collinear radiation relative to the event plane. This motivates defining a new object: the \textbf{C-shoulder jet function}.

\paragraph{Definition.}
The C-shoulder jet function for a quark jet is defined as
\begin{equation}
\label{eq:JC-def}
J^C_q(m^2,\mu) = \frac{1}{2N_c}\,{\rm tr}\,\langle 0 | \bar{\chi}_n\, \delta(m^2 - \widehat{M}^C_J)\,\frac{\not{\!\bar{n}}}{2}\, \chi_n | 0 \rangle\,,
\end{equation}
where $\chi_n$ is the gauge-invariant collinear quark field in SCET, and the C-parameter measurement operator is
\begin{equation}
\label{eq:MC-def}
\widehat{M}^C_J = 12\sum_{a} \frac{(\widehat{p}_{a\perp}^x)^2}{\bar{n}\cdot \widehat{p}_a}\,.
\end{equation}
Here $p_{a\perp}^x$ is the component of the transverse momentum perpendicular to the event plane, and the sum runs over all collinear particles in the jet. The factor of 12 is the geometric factor $\alpha_i$ from the Mercedes configuration. For a gluon jet:
\begin{equation}
J^C_g(m^2,\mu) = \frac{\omega}{2(N_c^2-1)}\,\langle 0 | \mathcal{B}_{n\perp}^\mu\, \delta(m^2 - \widehat{M}^C_J)\, \mathcal{B}_{n\perp\mu} | 0 \rangle\,.
\end{equation}

For a single collinear emission with invariant mass $m^2$ and azimuthal angle $\phi$, we have $(p_\perp^x)^2/(\bar{n}\cdot p) = m^2\sin^2\phi$, so $\widehat{M}^C_J \to 12 m^2\sin^2\phi$, recovering the measurement from Section~\ref{sec:factorization}.

\paragraph{One-loop result.}
At one loop, the C-shoulder jet function is computed by integrating the inclusive jet function over the azimuthal angle with the C-parameter measurement constraint. The calculation is presented in Appendix~\ref{app:Cjet}. Expanding the cusp term and absorbing the $\ln 3$ into the non-cusp anomalous dimension, the result is
\begin{equation}
\label{eq:JC-oneloop}
\boxed{
J^C_i(m^2,\mu) = \delta(m^2) + \frac{\as}{4\pi}\left[C_i\Gamma_0\left[\frac{\ln(m^2/\mu^2)}{m^2}\right]_+ + \gamma^C_{J,i}\left[\frac{1}{m^2}\right]_+ + \left(c_J^{C,i} - \frac{C_i\Gamma_0\pi^2}{12}\right)\delta(m^2)\right] + \order{\as^2},
}
\end{equation}
where $C_q = \CF$, $C_g = \CA$, $\Gamma_0 = 4$. The C-shoulder jet anomalous dimensions are
\begin{equation}
\label{eq:gamma-JC}
\gamma^C_{J,i} = \gamma_{J,i} - C_i\Gamma_0\ln 3\,,
\end{equation}
where $\gamma_{J,q} = -3\CF$ and $\gamma_{J,g} = -\beta_0$ are the inclusive jet anomalous dimensions. Explicitly:
\begin{equation}
\gamma^C_{J,q} = -3\CF - \Gamma_0\CF\ln 3 = -3\CF - 4\CF\ln 3\,, \qquad \gamma^C_{J,g} = -\beta_0 - \Gamma_0\CA\ln 3 = -\beta_0 - 4\CA\ln 3\,.
\end{equation}
The one-loop constants are shifted from the inclusive jet function:
\begin{equation}
\label{eq:cJCq}
c_J^{C,i} = c_J^i - \gamma_{J,i}\ln 3 + \frac{C_i\Gamma_0}{2}\left(\ln^2 3 + \frac{\pi^2}{3}\right),
\end{equation}
where the inclusive jet constants are
\begin{equation}
\label{eq:cJq}
c_J^q = \CF\left(7 - \frac{2\pi^2}{3}\right), \qquad
c_J^g = \CA\left(\frac{67}{9} - \frac{2\pi^2}{3}\right) - \frac{20}{9}\TF\nf\,.
\end{equation}

The $\ln 3$ shift in the anomalous dimension arises from averaging $\ln\sin^2\phi$ over the azimuthal angle: $\langle\ln\sin^2\phi\rangle = -2\ln 2$, combined with the geometric factor $\ln 12 - 2\ln 2 = \ln 3$. This structure parallels the soft function, where the Mercedes geometry also produces a $\ln 3$ in the anomalous dimension.

\subsection{Soft function}

The soft function $S(k,\mu)$ was defined in Eq.~\eqref{eq:soft-def} as the vacuum matrix element of soft Wilson lines with the C-parameter measurement operator $\widehat{M}_S = 4\sum_a k_{a,\perp}^2/k_a^0$. The argument $k$ has mass dimension one and scales as $k \sim Qc$.

At one loop, the soft function integral takes the form
\begin{equation}
S^{(1)}(k) = \sum_{i<j}(\mathbf{T}_i \cdot \mathbf{T}_j)\int\!\frac{d^dp}{(2\pi)^{d-1}} \frac{n_i \cdot n_j}{(n_i \cdot p)(n_j \cdot p)}\,\delta(p^2)\,\theta(p^0)\,\delta\!\left(k - \frac{4p_\perp^2}{p^0}\right).
\end{equation}
After performing the color algebra using $\mathbf{T}_1 + \mathbf{T}_2 + \mathbf{T}_3 = 0$, the color factor is $\mathcal{C} = 2\CF + \CA$, reflecting soft gluon coupling to two quark lines (each $\CF$) and one gluon line ($\CA$). The Wilson line directions at 120$^\circ$ angles give $n_i \cdot n_j = 3$, which introduces geometric factors in the angular integration.

The one-loop soft function is computed in Appendix~\ref{app:soft-direct} by reducing the angular master integral to a one-dimensional integral with a closed-form hypergeometric representation. The renormalized soft function is
\begin{equation}
\label{eq:soft-result}
\boxed{
S(k,\mu) = \delta(k) + \frac{\as(\mu)}{4\pi}\left[
-\mathcal{C}\Gamma_0\left[\frac{\ln(k/\mu)}{k}\right]_+ + 2\gamma_S\left[\frac{1}{k}\right]_+ + \left(c_S^1 + \frac{\mathcal{C}\pi^2}{3}\right)\delta(k)
\right] + \order{\as^2}
}
\end{equation}
where $\Gamma_0 = 4$ is the cusp anomalous dimension. The factor of 2 in the coefficient $2\gamma_S$ of $[1/k]_+$ arises because the soft function argument $k$ has mass dimension 1 (in contrast to the jet function argument $m^2$ which has dimension 2). The soft non-cusp anomalous dimension is
\begin{equation}
\gamma_S = 2\mathcal{C}\ln 3\,.
\end{equation}
The $\ln 3$ reflects the Mercedes geometry ($n_i \cdot n_j = 3$). This structure parallels the C-shoulder jet function, with the $\ln 3$ absorbed into the non-cusp anomalous dimension rather than appearing as a shift in the cusp logarithm.

The Laplace-space matching constant is
\begin{equation}
\label{eq:c-delta-result}
c_S^1 = \mathcal{C}\left(-\frac{2\pi^2}{3} + 4\ln^2 2 - \frac{1}{3}\ln 27\ln 48 + 2\,{\rm Li}_2\!\left(-\frac{1}{3}\right)\right).
\end{equation}

\subsection{NLO prediction from SCET}
\label{sec:SCET-NLO-prediction}

We now derive the NLO singular structure by expanding the factorization formula Eq.~\eqref{eq:master-formula} to $\order{\as^2}$. This provides a cross-check of the SCET framework and predictions that can be validated against fixed-order calculations.

The kernel $K(c,\mu)$ defined in Eq.~\eqref{eq:kernel-convolution} is the convolution of the hard, jet, and soft functions. At tree level, all functions reduce to delta functions, giving $K^{(0)}(c) = \delta(c)$. At one loop, we expand $K = H \cdot (J^C_q \otimes J^C_{\bar{q}} \otimes J^C_g \otimes S)$ to $\order{\as}$.

The one-loop hard function is a multiplicative prefactor: $H = 1 + (\as/4\pi)c_H^1 + \order{\as^2}$ when evaluated at $\mu = Q$.

The jet functions are naturally expressed in terms of the invariant mass variable $m^2$. Summing over the three jets at one loop:
\begin{equation}
J^C(m^2) \equiv J^C_q \otimes J^C_{\bar{q}} \otimes J^C_g = \delta(m^2) + \frac{\as}{4\pi}\left[\mathcal{C}\Gamma_0\left[\frac{\ln(m^2/\mu^2)}{m^2}\right]_+ + \gamma^C_J\left[\frac{1}{m^2}\right]_+ + \left(c_J^{C,{\rm tot}} - \frac{\mathcal{C}\Gamma_0\pi^2}{12}\right)\delta(m^2)\right],
\end{equation}
where $\mathcal{C} = 2\CF + \CA$, $\gamma^C_J = 2\gamma^C_{J,q} + \gamma^C_{J,g} = -6\CF - \beta_0 - \Gamma_0\mathcal{C}\ln 3 = -6\CF - \beta_0 - 4\mathcal{C}\ln 3$ is the total C-shoulder jet anomalous dimension, and $c_J^{C,{\rm tot}} = 2c_J^{C,q} + c_J^{C,g}$.

The soft function is expressed in terms of the variable $k$ with mass dimension one:
\begin{equation}
S(k) = \delta(k) + \frac{\as}{4\pi}\left[-\mathcal{C}\Gamma_0\left[\frac{\ln(k/\mu)}{k}\right]_+ + 2\gamma_S\left[\frac{1}{k}\right]_+ + \left(c_S^1 + \frac{\mathcal{C}\pi^2}{3}\right)\delta(k)\right],
\end{equation}
where $\gamma_S = 2\mathcal{C}\ln 3$.

To form $K(c)$, we convolve the jet and soft functions with the constraint that relates their arguments to $c$:
\begin{equation}
K(c) = H \int\! dm^2\, dk\; J^C(m^2)\, S(k)\, \delta(cQ^2 - m^2 - kQ)\,.
\end{equation}
At NLO, only one of $J^C$ or $S$ contributes a non-trivial distribution while the other is $\delta(m^2)$ or $\delta(k)$. Setting $\mu = Q$:
\begin{equation}
\label{eq:K-oneloop}
K(c) = \delta(c) + \frac{\as}{4\pi}\left[
-2\mathcal{C}\Gamma_0\left[\frac{\ln c}{c}\right]_+ + (\gamma_J^{C} + 2\gamma_S)\left[\frac{1}{c}\right]_+ + \Delta^1\,\delta(c)
\right] + \order{\as^2}\,,
\end{equation}
where $\gamma_J^C$ and $\gamma_S$ are the non-cusp anomalous dimensions of the C-shoulder jet and soft functions. The coefficient of $[\ln c/c]_+$ is $-2\Gamma_0\mathcal{C}$ from the cusp anomalous dimension (factor of 2 from jets plus soft). The factor of 2 multiplying $\gamma_S$ arises from the different mass dimensions of the jet and soft function arguments: the jet function depends on $m^2$ (dimension 2) while the soft function depends on $k$ (dimension 1), leading to different coefficients in the Laplace-space matching functions. The C-shoulder jet anomalous dimension is $\gamma_J^{C} = 2\gamma_{J,q}^{C} + \gamma_{J,g}^{C} = -6\CF - \beta_0 - \mathcal{C}\Gamma_0\ln 3 = -6\CF - \beta_0 - 4\mathcal{C}\ln 3$ (see Appendix~\ref{app:Cjet}), while the soft anomalous dimension is $\gamma_S = 2\mathcal{C}\ln 3$ (see Appendix~\ref{app:soft-direct}). Therefore:
\begin{equation}
\gamma_J^{C} + 2\gamma_S = (-6\CF - \beta_0 - 4\mathcal{C}\ln 3) + 2(2\mathcal{C}\ln 3) = -6\CF - \beta_0\,.
\end{equation}
The $\ln 3$ terms cancel: the $-4\mathcal{C}\ln 3$ from the C-shoulder jet functions is exactly cancelled by $+4\mathcal{C}\ln 3$ from the soft function. The constant term is
\begin{equation}
\label{eq:Delta1}
\Delta^1 = c_H^1 + 2c_J^{C,q} + c_J^{C,g} + c_S^1\,,
\end{equation}
with $c_H^1$ from Eq.~\eqref{eq:cH1}, $c_J^{C,i}$ from Eq.~\eqref{eq:cJCq}, and $c_S^1$ from Eq.~\eqref{eq:c-delta-result}.

The cumulant $R(c)$ is defined in Eq.~\eqref{eq:cumulant-R} as the integral of the kernel. Using
\begin{equation}
\int_0^c \left[\frac{\ln c'}{c'}\right]_+ dc' = \frac{1}{2}\ln^2 c\,, \qquad
\int_0^c \left[\frac{1}{c'}\right]_+ dc' = \ln c\,,
\end{equation}
we obtain
\begin{equation}
\label{eq:R-oneloop}
R(c) = 1 + \frac{\as}{4\pi}\left[
-\mathcal{C}\Gamma_0\ln^2 c + (\gamma_J^{C} + 2\gamma_S)\ln c + \Delta^1
\right] + \order{\as^2}\,.
\end{equation}

Substituting the cumulant Eq.~\eqref{eq:R-oneloop} into the factorization theorem Eq.~\eqref{eq:master-formula}:
\begin{equation}
\frac{1}{\sigma_0}\frac{d\sigma}{dc}\bigg|_{\rm sing} = \frac{\as}{2\pi}\,A_{\rm Born}\!\left(\frac{3}{4}\right) \big(1 - R(c)\big)\,.
\end{equation}
Expanding to $\order{\as^2}$:
\begin{equation}
\label{eq:SCET-full-expansion}
\frac{1}{\sigma_0}\frac{d\sigma}{dc}\bigg|_{\rm sing} = \left(\frac{\as}{2\pi}\right)^2 A_{\rm Born}\!\left(\frac{3}{4}\right) \left[\frac{\mathcal{C}\Gamma_0}{4}\ln^2 c - \frac{\gamma_J^{C} + 2\gamma_S}{2}\ln c - \frac{\Delta^1}{2}\right] + \order{\as^3}\,.
\end{equation}
The singular NLO distribution is therefore
\begin{equation}
\frac{1}{\sigma_0}\frac{d\sigma}{dc}\bigg|_{\rm sing}^{\rm NLO} = \left(\frac{\as}{2\pi}\right)^2 A_{\rm Born}\!\left(\frac{3}{4}\right) \left[\mathcal{C}\ln^2 c + B_1\ln c\right],
\end{equation}
where the double-log coefficient is the total cusp color factor
\begin{equation}
\label{eq:A2-coeff}
A_2 = \mathcal{C} = 2\CF + \CA\,,
\end{equation}
and the single-log coefficient is
\begin{equation}
\label{eq:B1-coeff}
B_1 = -\frac{\gamma_J^{C} + 2\gamma_S}{2} = 3\CF + \frac{\beta_0}{2}\,.
\end{equation}
This matches the Catani--Webber result~\cite{Catani:1997xc}.

\subsection{Validation of the singular distribution}
\label{sec:singular-validation}

\begin{figure}[t]
\centering
\includegraphics[width=\textwidth]{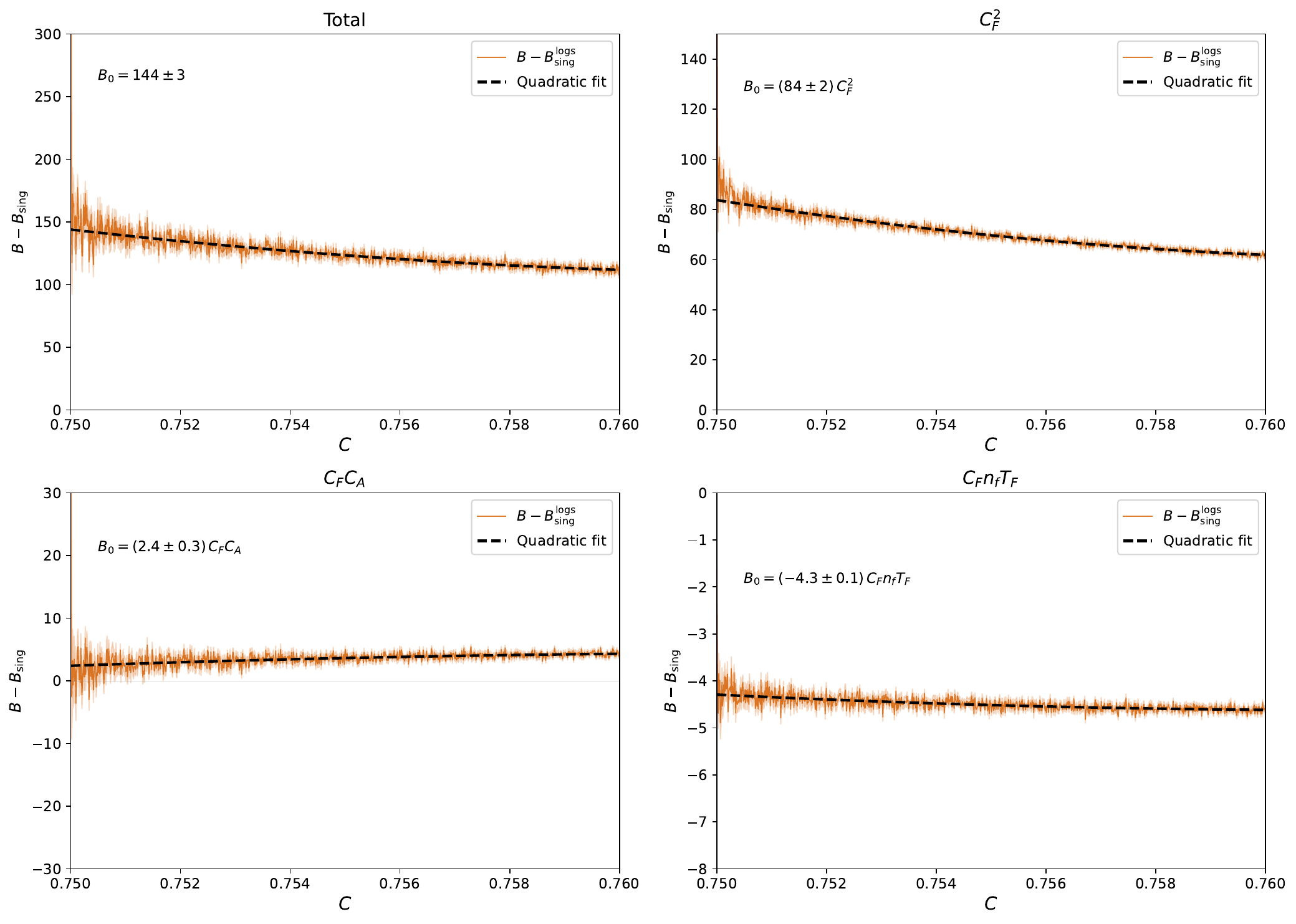}
\caption{Non-singular distribution $B(c) - B_{\rm sing}^{\rm logs}(c)$ for each color channel in the shoulder region, where $B_{\rm sing}^{\rm logs}$ contains only the $\ln c$ and $\ln^2 c$ terms predicted by SCET. The smooth extrapolation to $c \to 0$ validates the singular coefficients. Quadratic fits (dashed) give the non-singular intercepts $B_0$ for each channel.}
\label{fig:log-subtraction}
\end{figure}

The SCET prediction for the singular distribution can be validated against fixed-order EVENT2 calculations. Writing $c = \frac{8}{3}(C - \frac{3}{4})$ and $\beta_0 = \frac{11}{3}\CA - \frac{4}{3}\nf\TF$, the singular terms decompose by color structure:
\begin{align}
\label{eq:Bsing-logs}
B_{\rm sing}^{\rm logs}(c) &= A\Big(\frac{3}{4}\Big)\Big[\CF^2\big(2\ln^2 c + 3\ln c\big) + \CF\CA\big(\ln^2 c + \tfrac{11}{6}\ln c\big) - \CF\nf\TF\,\tfrac{2}{3}\ln c\Big]\,.
\end{align}
Figure~\ref{fig:log-subtraction} shows the result of subtracting these logarithmic terms from the EVENT2 data for each color channel. The smooth extrapolation to $c \to 0$ confirms that the singular distribution agrees with both Catani--Webber and EVENT2, validating the SCET calculation. The non-singular intercept is
\begin{equation}
\label{eq:B0-intercepts}
B_0 = \CF^2\,B_0^{\CF^2} + \CF\CA\,B_0^{\CF\CA} + \CF\nf\TF\,B_0^{\CF\nf\TF} = 144 \pm 3\,,
\end{equation}
with channel coefficients
\begin{equation}
B_0^{\CF^2} = 84 \pm 2\,,\qquad B_0^{\CF\CA} = 2.4 \pm 0.3\,,\qquad B_0^{\CF\nf\TF} = -4.3 \pm 0.1\,.
\end{equation}
The intercept $B_0$ represents the size of the non-singular contributions to the C-parameter distribution in the shoulder region. As discussed in Section~\ref{sec:resummed-xsec}, the integral $\int_0^\infty K(c')\,dc'$ appearing in the factorization formula is formally divergent, and this $c$-independent divergence is absorbed into the non-singular matching. Consequently, $B_0$ cannot be predicted within SCET and must be extracted from fixed-order calculations. This intercept does not play any special role in the final result---for matching we use the full non-singular distribution $B_{\rm NS}(c)$ rather than just its $c \to 0$ limit.


\section{NLL Resummation}
\label{sec:resummation}

We now derive the resummed NLL formula for the C-parameter distribution across the shoulder. The basic idea is that the factorization theorem Eq.~\eqref{eq:master-formula} expresses the cross section in terms of hard, jet, and soft functions that each satisfy renormalization group equations. Solving these RG equations and evolving the functions from their canonical scales (where they contain no large logarithms) to a common scale resums the Sudakov logarithms to all orders. The perturbative ingredients required for NLL accuracy---the two-loop cusp and one-loop non-cusp anomalous dimensions, two-loop beta function, and running coupling---were assembled in Section~\ref{sec:ingredients}.

\subsection{The resummed cumulant}

The cumulant $R(c)$ is the integral of the singular kernel from the factorization theorem. The momentum-space formula is
\begin{equation}
\label{eq:R-integral}
R(c) = H(\mu)\int_0^\infty\! dm^2\, J^C(m^2,\mu) \int_0^\infty\! dk\, S(k,\mu)\, \theta\!\left(Q^2 c - m^2 - Qk\right),
\end{equation}
where the theta function implements the constraint that the total collinear and soft contributions do not exceed $c$. The combined jet function $J^C(m^2,\mu)$ is the convolution of the three individual jet functions:
\begin{equation}
J^C(m^2,\mu) = \int_0^{m^2}\! dm_1^2 \int_0^{m^2-m_1^2}\! dm_2^2\, J_q^C(m_1^2,\mu)\, J_{\bar{q}}^C(m_2^2,\mu)\, J_g^C(m^2 - m_1^2 - m_2^2,\mu)\,.
\end{equation}

At tree level $R(c) = \theta(c)$, giving a step function. Beyond tree level, $R(c)$ receives radiative corrections that smooth out the step discontinuity. The key feature is that $R(c) \to 0$ as $c \to 0^+$ due to Sudakov suppression from the running between scales.

It is convenient to work in Laplace space, where the convolutions become products and the RG evolution can be solved analytically. The resummed cumulant takes the form
\begin{equation}
\label{eq:R-derivative}
R(c) = \Pi(\mu_H, \mu_J, \mu_S)\; h(L_H)\; \tilde{j}_q\big(\partial_\eta + L_R\big)^2\, \tilde{j}_g\big(\partial_\eta + L_R\big)\, \tilde{s}(\partial_\eta)\; \mathcal{F}(\eta)\,,
\end{equation}
where the derivative operators $\partial_\eta$ arise because logarithms of $c$ in Laplace space translate to derivatives acting on the base function, $L_H = \ln(Q^2/\mu_H^2)$, and $L_R = \ln(Q\mu_S/\mu_J^2)$.

The evolution kernel $\Pi$ collects all RG evolution factors:
\begin{equation}
\label{eq:Pi-def}
\Pi(\mu_H, \mu_J, \mu_S) = \exp\Big[2\mathcal{C}\big(S(\mu_H, \mu_J) + S(\mu_S, \mu_J)\big) - 2A_{\gamma_H}(\mu_H,\mu_J) + 2A_{\gamma_S}(\mu_S,\mu_J)\Big] \left(\frac{Q^2}{\mu_H^2}\right)^{\!-\mathcal{C}\,A_\Gamma(\mu_H, \mu_J)},
\end{equation}
where $\mathcal{C} = 2\CF + \CA = 17/3$ is the total cusp color factor from two quark jets plus one gluon jet. The Sudakov integral $S(\nu,\mu)$ and evolution integrals $A_\Gamma$, $A_{\gamma_X}$ are defined as
\begin{align}
S(\nu,\mu) &= -\int_{\as(\nu)}^{\as(\mu)}\! \frac{d\as}{\beta(\as)}\, \Gamma_{\rm cusp}(\as) \int_{\as(\nu)}^{\as}\! \frac{d\as'}{\beta(\as')}\,, \\
A_\Gamma(\nu,\mu) &= -\int_{\as(\nu)}^{\as(\mu)}\! \frac{d\as}{\beta(\as)}\, \Gamma_{\rm cusp}(\as)\,, \\
A_{\gamma_X}(\nu,\mu) &= -\int_{\as(\nu)}^{\as(\mu)}\! \frac{d\as}{\beta(\as)}\, \gamma_X(\as)\,,
\end{align}
with explicit NLL formulas collected in Appendix~\ref{app:NLL}.

The matching functions $h$, $\tilde{j}_i$, $\tilde{s}$ encode the fixed-order corrections at each scale. They have the general 1-loop structure
\begin{equation}
\tilde{j}^C(L) = 1 + \frac{\as}{4\pi}\left[\mathcal{C}\frac{\Gamma_0}{2}L^2 + \gamma_J^{C} L + c_J^{C}\right],
\end{equation}
where $\gamma_J^C$ and $c_J^C$ are the combined one-loop anomalous dimension and constant. The hard and soft matching functions $h(L_H)$ and $\tilde{s}(L)$ have similar structure; explicit expressions are given in Appendix~\ref{app:NLL}.

The base function $\mathcal{F}(\eta)$ encodes the Sudakov suppression. With the Sudakov exponent
\begin{equation}
\eta = 2\mathcal{C}\,A_\Gamma(\mu_J, \mu_S)\,,
\end{equation}
the base function is
\begin{equation}
\mathcal{F}(\eta) = \frac{e^{-\gamma_E \eta}}{\Gamma(1+\eta)}\left(\frac{c Q}{\mu_S}\right)^\eta,
\end{equation}
where $\gamma_E$ is the Euler-Mascheroni constant. At small $c$ with canonical scales $\mu_S \sim cQ$, the exponent $\eta \to \infty$ and $\mathcal{F}(\eta) \to 0$, so $R(c) \to 0$. This Sudakov suppression is the origin of the smooth shoulder.

Scale-independence of the factorized cross section requires the anomalous dimensions to satisfy
\begin{equation}
\label{eq:RG-consistency}
\gamma_H = \gamma_J^{C} + \gamma_S\,.
\end{equation}
This relation must hold order by order in $\as$. For the one-loop coefficients computed in Section~\ref{sec:ingredients}, $\gamma_H^{(0)} = -6\CF - \beta_0 - 2\mathcal{C}\ln 3$, $\gamma_{J}^{C,(0)} = -6\CF - \beta_0 - 4\mathcal{C}\ln 3$, and $\gamma_S^{(0)} = 2\mathcal{C}\ln 3$, which indeed satisfy $\gamma_H^{(0)} = \gamma_J^{C,(0)} + \gamma_S^{(0)}$.

The canonical scale choices that minimize logarithms in the matching functions are
\begin{equation}
\label{eq:canonical-scales}
\mu_H = Q\,, \qquad \mu_J = Q\sqrt{c}\,, \qquad \mu_S = cQ\,,
\end{equation}
where $c = (8/3)(C - 3/4)$ is the rescaled shoulder variable absorbing the geometric factor from the Mercedes configuration. With these choices, the hard function logarithm $L_H = \ln(Q^2/\mu_H^2) = 0$, while the jet and soft logarithms are traded for the exponent $\eta$ via the RG evolution.

As a consistency check, expanding $R(c)$ to $\order{\as}$ reproduces the singular structure derived in Section~\ref{sec:SCET-NLO-prediction}:
\begin{equation}
R(c) = 1 + \frac{\as}{4\pi}\Big[-\mathcal{C}\Gamma_0\ln^2 c + (\gamma_J^C + 2\gamma_S)\ln c + \Delta^1 + \frac{\mathcal{C}\Gamma_0\pi^2}{12}\Big] + \order{\as^2}\,,
\end{equation}
where $\Delta^1 = c_H^1 + c_J^{C} + c_S^1$ is the sum of matching constants defined in Eq.~\eqref{eq:Delta1}. The additional $\pi^2$ term arises from the $k_2\,\partial_\eta^2 \mathcal{F}$ contribution in the kernel expansion, using $\partial_\eta^2 \mathcal{F}|_{\eta\to 0} = -\pi^2/6$.
Substituting into $d\sigma/dc \propto -R(c)$ gives the singular NLO coefficients $A_2 = \mathcal{C}$ and $B_1 = 3\CF + \beta_0/2$, in agreement with Eqs.~\eqref{eq:A2-coeff} and \eqref{eq:B1-coeff}.

\subsection{Matching to fixed order}
\label{sec:matching-theory}

To obtain NLO-accurate predictions, the resummed distribution must be matched to the full fixed-order result. This requires understanding the singular distribution at different levels of accuracy.

Above the shoulder, the NLO coefficient $B(C)$ can be decomposed into singular and non-singular pieces. There are three natural levels of the singular distribution, distinguished by which constant terms are included. With the rescaled variable $c = (8/3)(C - 3/4)$, the singular distribution $B_{\rm sing}^{\rm logs}(c)$ from Eq.~\eqref{eq:Bsing-logs} can be written as
\begin{align}
\label{eq:Bsing-hierarchy}
B_{\rm sing}^{\rm logs}(c) &= A_{3/4}\Big[(2\CF + \CA)\ln^2 c + (3\CF + \tfrac{\beta_0}{2}) \ln c\Big]\,, \\
B_{\rm sing}^{\rm NLL}(c) &= B_{\rm sing}^{\rm logs}(c) - \frac{\pi^2}{6}(2\CF + \CA)\, A_{3/4}\,, \nonumber\\
B_{\rm sing}^{\rm NLO}(c) &= B_{\rm sing}^{\rm logs}(c) - \frac{\pi^2}{6}(2\CF + \CA)\, A_{3/4} + \frac{c_H^1 + c_J^C + c_S^1}{2}\, A_{3/4}\,, \nonumber
\end{align}
where $A_{3/4} = A(3/4) = 256\pi\sqrt{3}\CF/243$. The first line contains only the logarithms of $c$, which when expanded in terms of $(C - 3/4)$ give
\begin{equation}
\ln c = \ln\tfrac{8}{3} + \ln (C-\tfrac{3}{4})\,, \qquad \ln^2 c = \ln^2\tfrac{8}{3} + 2\ln\tfrac{8}{3}\ln (C-\tfrac{3}{4}) + \ln^2(C-\tfrac{3}{4})\,,
\end{equation}
generating both logarithms and constants from the rescaling. The second line adds the $\pi^2$ correction from the inverse Laplace transform of $L^2$ terms. The third line adds the one-loop matching constants for full NLO accuracy.

\begin{figure}[t]
\centering
\includegraphics[width=0.7\textwidth]{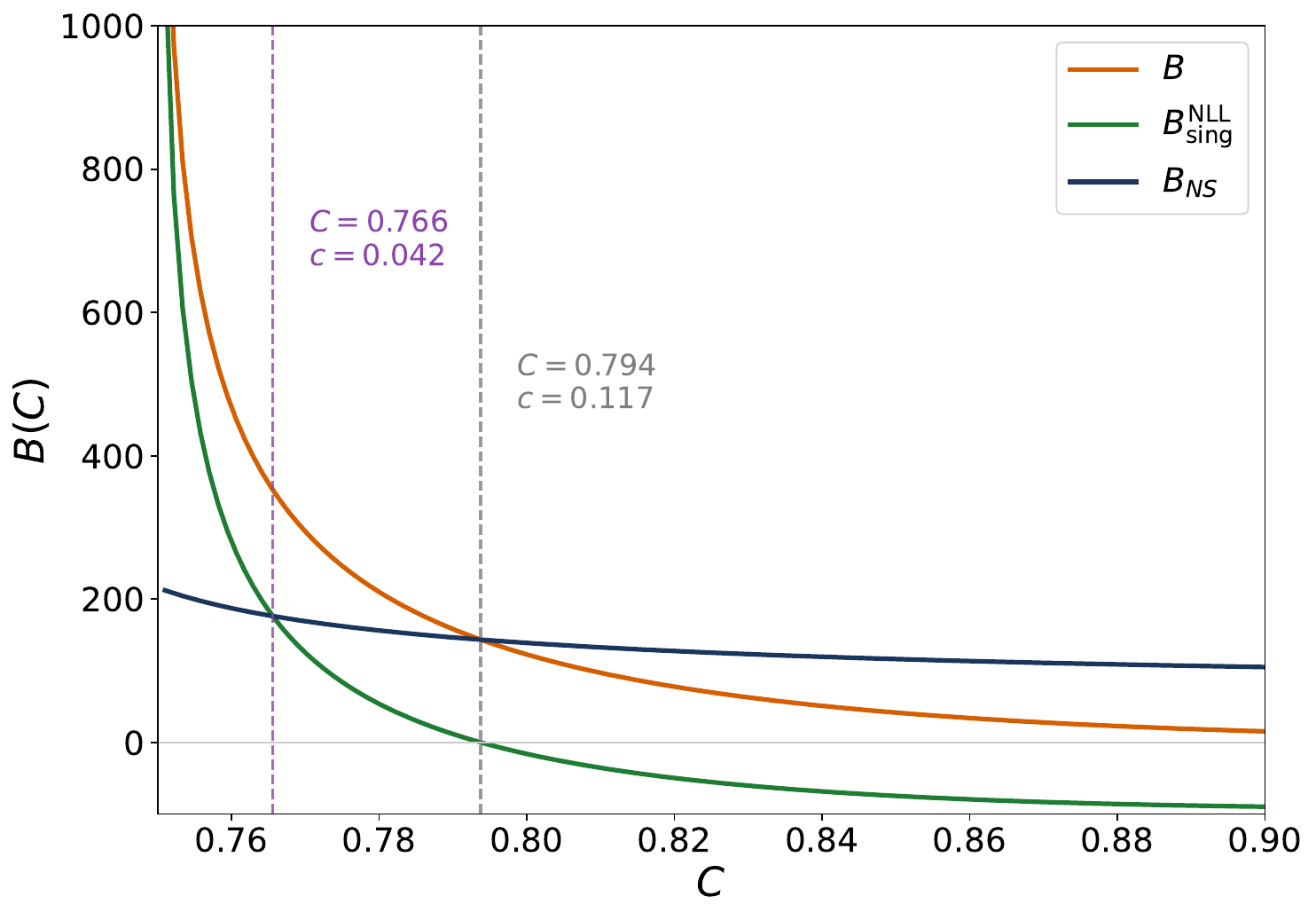}
\caption{Comparison of the NLO distribution $B(C)$ (red), the NLL singular prediction $B_{\rm sing}^{\rm NLL}(C)$ (green), and the non-singular distribution $B_{\rm NS}(C) = B(C) - B_{\rm sing}^{\rm NLL}(C)$ (blue) in the shoulder region, where $c = (8/3)(C - 3/4)$. The singular coefficients are fixed by SCET (see Eq.~\eqref{eq:SCET-full-expansion}). The crossing points where $B_{\rm sing}^{\rm NLL} = 0$ and where $B_{\rm sing}^{\rm NLL} = B_{\rm NS}$ are indicated; beyond the latter, non-singular corrections dominate over resummed contributions.}
\label{fig:B-crossings}
\end{figure}

Figure~\ref{fig:B-crossings} shows the decomposition $B = B_{\rm sing}^{\rm NLL} + B_{\rm NS}$, where $B_{\rm sing}^{\rm NLL}$ crosses zero at $C \approx 0.80$ and equals $B_{\rm NS}$ at $C \approx 0.77$. The crossing points indicate the scale at which resummation effects become comparable to non-singular corrections. The full NLO distribution $B(C)$ is well-described by
\begin{equation}
\label{eq:B-fit-formula}
B(C) = B_{\rm sing}^{\rm logs}(c) + \frac{B_0 + a_1 c}{1 + b_1 c + b_2 c^2}\,,
\end{equation}
where $B_0$ is given in Eq.~\eqref{eq:B0-intercepts} and $a_1 = 1834$, $b_1 = 22.4$, $b_2 = 100$.

The non-singular piece for NLL matching is defined as
\begin{equation}
B_{\rm NS}(c) = B(C) - B_{\rm sing}^{\rm NLL}(c)\,,
\end{equation}
which differs from Eq.~\eqref{eq:B-fit-formula} by the $\pi^2$ constant. The matched cross section is
\begin{equation}
\label{eq:matched-formula}
\frac{1}{\sigma_0}\frac{d\sigma}{dC} =
\begin{cases}
\displaystyle \frac{\as}{2\pi} A(C) + \left(\frac{\as}{2\pi}\right)^{\!2} B(C) & C < 3/4 \\[1.2em]
\displaystyle \frac{\as}{2\pi}\, A_{3/4}\, \big(1 - R(c)\big) + \left(\frac{\as}{2\pi}\right)^{\!2} B_{\rm NS}(c) & C > 3/4
\end{cases}
\end{equation}
As $c \to 0^+$, Sudakov suppression gives $R(c) \to 0$, so the $\order{\as}$ piece approaches $(\as/2\pi)A_{3/4}$ continuously from both sides. The $\order{\as^2}$ pieces need not match exactly at $C = 3/4$, but the discontinuity is small and phenomenologically harmless.


\section{Numerical Validation and Matching}
\label{sec:matching}

We now validate the SCET predictions against fixed-order calculations and present the matched NLL+NLO distribution with theoretical uncertainties.

\subsection{Validation of resummation}
\label{sec:validation-resummation}

The resummation formula Eq.~\eqref{eq:R-derivative} depends on the choice of renormalization scales $\mu_H$, $\mu_J$, and $\mu_S$. To validate that the resummed prediction correctly captures the singular structure, we compare the resummed distribution with the fixed-order singular distribution for different scale choices.

Consider a one-parameter family of scale choices parameterized by an exponent $t$:
\begin{equation}
\label{eq:t-scales}
\mu_H = Q\,, \qquad \mu_J = Q\, c^{t/2}\,, \qquad \mu_S = Q\, c^t\,.
\end{equation}
The canonical scales Eq.~\eqref{eq:canonical-scales} correspond to $t = 1$, while $t = 0$ gives fixed scales $\mu_H = \mu_J = \mu_S = Q$ with no resummation. For intermediate values of $t$, the scale hierarchy $\mu_S < \mu_J < \mu_H$ is reduced, giving less Sudakov suppression.

\begin{figure}[t]
\centering
\includegraphics[width=\textwidth]{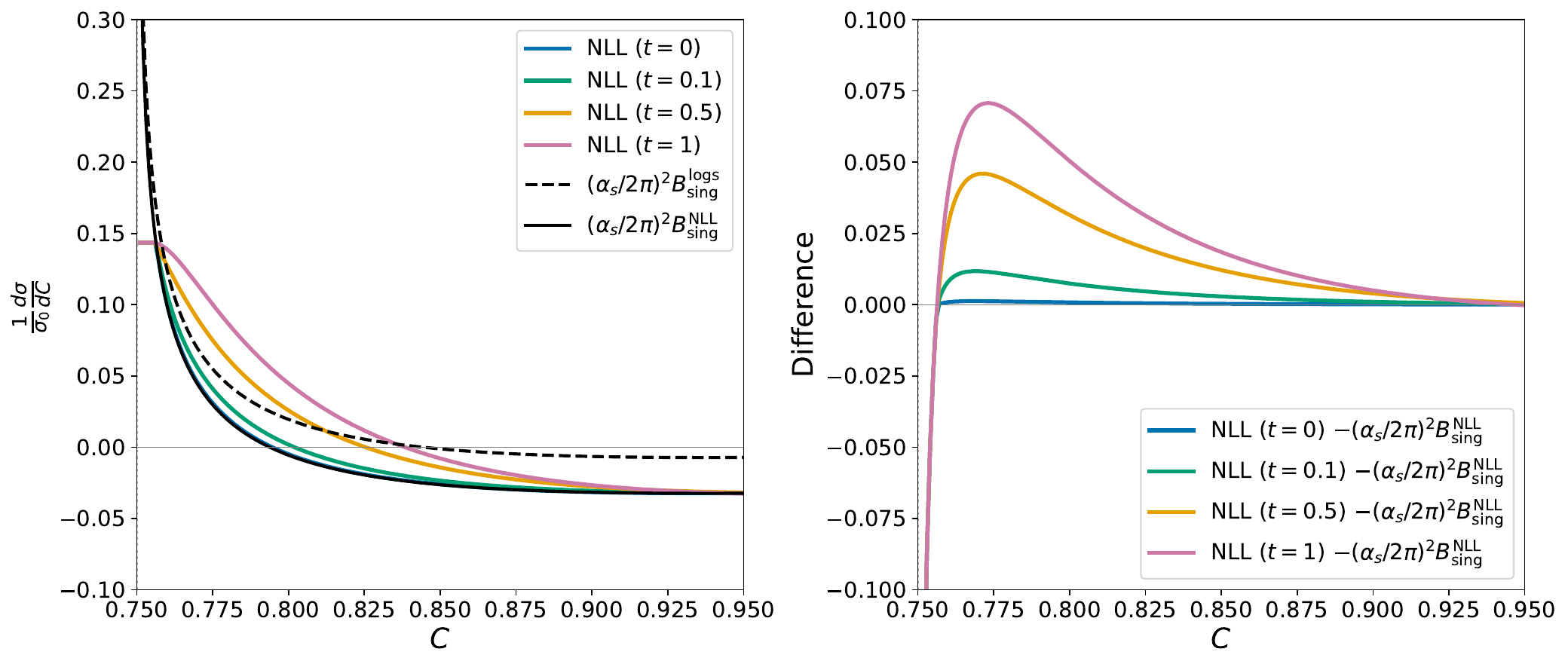}
\caption{Comparison of the resummed contribution $(\as/2\pi) A(3/4)(1 - R_{\rm NLL})$ with the singular distributions $(\as/2\pi)^2 B_{\rm sing}^{\rm logs}$ (dashed) and $(\as/2\pi)^2 B_{\rm sing}^{\rm NLL}$ (solid) for different scale exponents $t$ defined in Eq.~\eqref{eq:t-scales}. The canonical choice $t = 1$ gives maximal Sudakov suppression, while $t = 0$ corresponds to fixed scales with no resummation. For numerical stability, the $t = 0$ curve uses $t = 0.01$. Left: both contributions to the cross section. Right: difference from $B_{\rm sing}^{\rm NLL}$. The agreement at small $t$ validates that the resummation correctly reproduces the singular structure.}
\label{fig:NLL-tcurves}
\end{figure}

Figure~\ref{fig:NLL-tcurves} compares the resummed contribution $(\as/2\pi) A(3/4)(1 - R_{\rm NLL})$ with the fixed-order singular distribution $(\as/2\pi)^2 B_{\rm sing}^{\rm NLL}$ for different values of $t$. In the limit $t \to 0$ (fixed scales), the resummed contribution should approach the singular distribution, since the resummation reduces to a fixed-order expansion when all scales are equal. Indeed, the $t = 0$ curve closely tracks the $B_{\rm sing}^{\rm NLL}$ prediction, validating that the resummed formula correctly reproduces the singular structure at NLL accuracy.

As $t$ increases toward the canonical value $t = 1$, the Sudakov suppression becomes stronger, pushing the distribution away from the singular prediction. The difference between the resummed and singular distributions quantifies the higher-order effects captured by resummation. Near the shoulder at $C \approx 0.76$, the canonical $t = 1$ curve differs significantly from the singular distribution, reflecting the strong Sudakov suppression in this region.

\subsection{Profile scales}

The canonical scales Eq.~\eqref{eq:canonical-scales} are appropriate only for $c \ll 1$, where large logarithms $\ln c$ would otherwise appear in fixed-order perturbation theory. For larger $c$ approaching $c \sim 1$, the scales $\mu_J$ and $\mu_S$ become low enough that the hierarchy $\mu_S \ll \mu_J \ll \mu_H$ breaks down and the running coupling $\as(\mu_S)$ becomes unreliably large.

To obtain a well-behaved prediction across the full kinematic range, we employ profile functions~\cite{Ligeti:2008ac,Abbate:2010xh} that smoothly transition from canonical scales near the shoulder to fixed scales $\mu_H = \mu_J = \mu_S = Q$ far from the shoulder. The profile scales use the parameterization introduced in Section~\ref{sec:validation-resummation}:
\begin{equation}
\label{eq:profile-scales}
\mu_J(c) = Q\, c^{t_{\rm eff}(c)/2}\,, \qquad \mu_S(c) = Q\, c^{t_{\rm eff}(c)}\,,
\end{equation}
with the effective exponent
\begin{equation}
\label{eq:t-eff}
t_{\rm eff}(c) = t_{\rm min} + (1 - t_{\rm min}) \cdot \frac{1}{2}\left(1 - \tanh\frac{c - c_0}{w}\right).
\end{equation}
The profile function interpolates between $t_{\rm eff} = 1$ (canonical scales) for $c \ll c_0$ and $t_{\rm eff} = t_{\rm min}$ (nearly fixed scales) for $c \gg c_0$.

The central transition point $c_0$ is motivated by the crossing structure visible in Figure~\ref{fig:B-crossings}. Resummation is most important where the singular contribution $B_{\rm sing}$ dominates over the non-singular piece $B_{\rm NS}$. As shown in Figure~\ref{fig:B-crossings}, $B_{\rm sing} = B_{\rm NS}$ at $c \approx 0.04$ ($C \approx 0.77$), and $B_{\rm sing}$ crosses zero at $c \approx 0.12$ ($C \approx 0.79$). Beyond this point, the singular logarithms are no longer the dominant contribution and fixed-order perturbation theory becomes more appropriate. We therefore choose the central transition point $c_0 = 0.08$, roughly midway between these two crossings, corresponding to $C \approx 0.78$.

The central parameter values are:
\begin{itemize}
\item $c_0 = 0.08$: the transition point, corresponding to $C \approx 0.78$
\item $w = 0.05$: the transition width, chosen to give a smooth but localized transition
\item $t_{\rm min} = 0.01$: the minimum exponent, ensuring $\mu_S < \mu_J < \mu_H$ always holds
\end{itemize}

At $t_{\rm eff} = 1$, we recover the canonical scales with $\mu_J/\mu_S = c^{-1/2} \to \infty$ as $c \to 0$, giving maximal Sudakov suppression. As $t_{\rm eff} \to 0$, all scales converge to $Q$ and $R(c) \to 1$, so the cross section approaches fixed-order behavior.

\subsection{Results and uncertainties}

\begin{figure}[t]
\centering
\includegraphics[width=\textwidth]{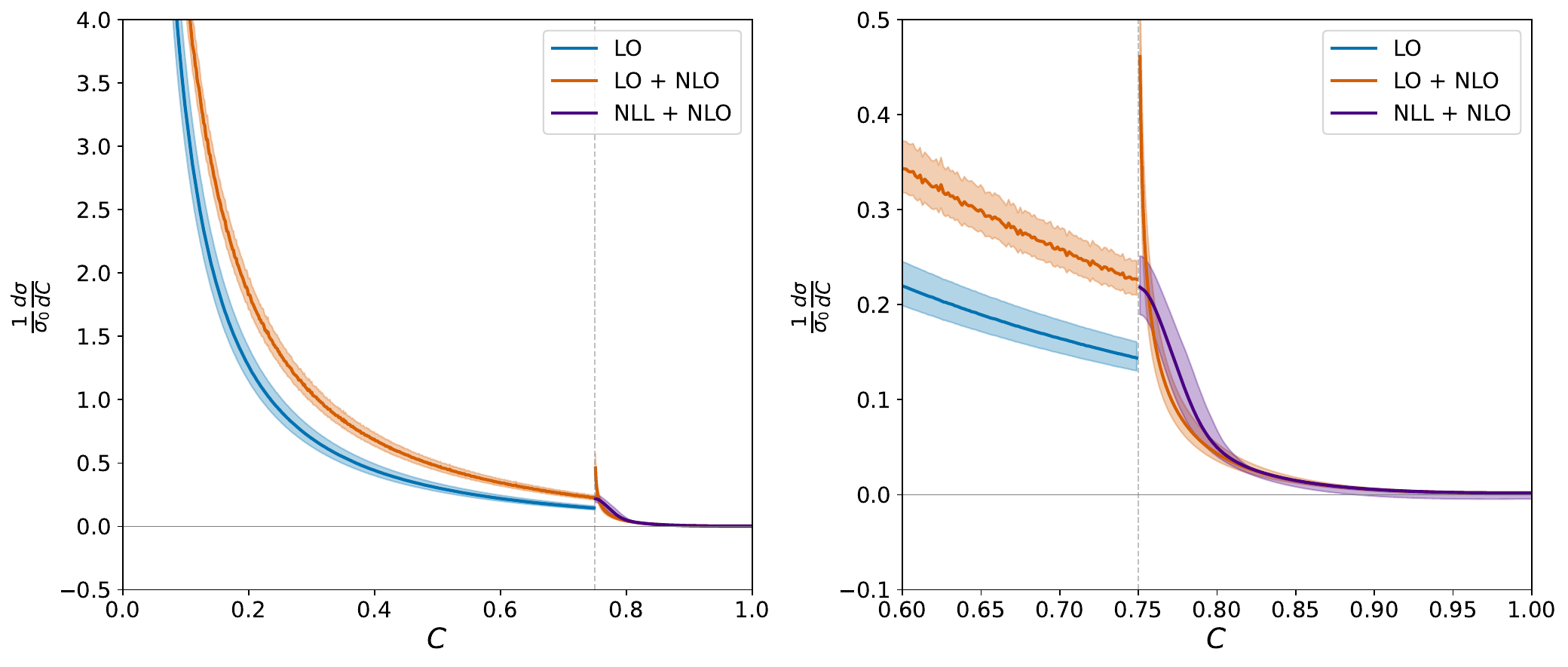}
\caption{C-parameter distribution at LO (blue), LO+NLO (orange), and NLL+NLO matched (purple) with $\as(M_Z) = 0.118$. The shaded bands show theoretical uncertainties from scale variations. Left: full distribution. Right: shoulder region $0.6 < C < 1.0$. The matched distribution smoothly crosses the shoulder, with the Sudakov factor ensuring $R(c) \to 0$ as $c \to 0$ so the NLL piece approaches $(\as/2\pi)A(3/4)$ from above.}
\label{fig:matched-distribution}
\end{figure}

Figure~\ref{fig:matched-distribution} shows the matched NLL+NLO distribution compared with LO and fixed-order LO+NLO. The resummation eliminates the unphysical logarithmic spike at $C = 3/4^+$ present in fixed-order NLO, producing a smooth Sudakov shoulder. The left panel shows the full C-parameter range, while the right panel focuses on the shoulder region $0.6 < C < 1.0$ where the resummation effects are most pronounced.

\paragraph{Fixed-order uncertainty bands.}
The LO and LO+NLO uncertainty bands in Figure~\ref{fig:matched-distribution} are obtained by varying the renormalization scale $\mu$ by a factor of two around the central value $\mu = Q$. The fixed-order cross section is
\begin{equation}
\frac{1}{\sigma_0}\frac{d\sigma}{dC} = \frac{\as(\mu)}{2\pi} A(C) + \left(\frac{\as(\mu)}{2\pi}\right)^{\!2} \left[B(C) + \frac{\beta_0}{2} A(C) \ln\frac{\mu^2}{Q^2}\right] + \order{\as^3}\,,
\end{equation}
where the $(\beta_0/2) \ln(\mu^2/Q^2)$ term arises from running $\as$ from $Q$ to $\mu$. The uncertainty band is obtained by:
\begin{itemize}
\item Renormalization scale: $\mu/Q \in [0.5, 2.0]$ (central: $1.0$)
\end{itemize}

\paragraph{NLL+NLO uncertainty band.}
The resummed cross section depends on three scales: the hard scale $\mu_H$, the jet scale $\mu_J$, and the soft scale $\mu_S$. Rather than varying these independently, we parameterize them through the profile function Eq.~\eqref{eq:profile-scales}, which ensures the correct limiting behavior: canonical scales $\mu_J = Q\sqrt{c}$, $\mu_S = Qc$ near the shoulder where resummation is important, transitioning to fixed scales $\mu_J = \mu_S = \mu_H$ far from the shoulder where fixed-order perturbation theory applies.

The profile function is controlled by the effective exponent $t_{\rm eff}(c)$ in Eq.~\eqref{eq:t-eff}, which interpolates between $t_{\rm eff} = 1$ (canonical) and $t_{\rm eff} = t_{\rm min}$ (nearly fixed). Varying the profile parameters effectively varies all three scales in a correlated way that respects the scale hierarchy $\mu_S < \mu_J < \mu_H$. The uncertainty band is constructed by varying:
\begin{itemize}
\item Hard scale: $\mu_H/Q \in [0.5, 2.0]$ (central: $1.0$)
\item Jet and soft scales via $t_{\rm min} \in [0.005, 0.02]$ (central: $0.01$)
\item Transition point: $c_0 \in [0.03, 0.13]$ (central: $0.08$)
\item Transition width: $w \in [0.03, 0.07]$ (central: $0.05$)
\end{itemize}
Varying $t_{\rm min}$ directly changes $t_{\rm eff}$ and thus the jet and soft scales through $\mu_J = Qc^{t_{\rm eff}/2}$ and $\mu_S = Qc^{t_{\rm eff}}$. Varying $c_0$ and $w$ changes where and how sharply the scales transition from canonical to fixed, which also affects $\mu_J$ and $\mu_S$ in the transition region. This correlated variation subsumes independent variations of $\mu_J$ and $\mu_S$ while maintaining the required hierarchy.

The uncertainty band is constructed by computing the cross section on a grid of parameter values (11 values for $c_0$, 3 for $w$, 3 for $t_{\rm min}$, and 5 for $\mu_H$), then taking the envelope of all valid combinations. The resulting uncertainty is approximately $\pm 15$--$20\%$ in the shoulder region $0.75 < C < 0.85$, comparable to other NLL event shape predictions~\cite{Becher:2008cf,Abbate:2010xh}. The uncertainty is largest near the shoulder at $C = 3/4$ where the resummation effects are strongest, and decreases for larger $C$ where fixed-order perturbation theory dominates.

Note that the distribution is not continuous at $C = 3/4$: while the $\order{\as}$ piece connects continuously as discussed in Sections~\ref{sec:step-discontinuity} and~\ref{sec:matching-theory}, the $\order{\as^2}$ pieces do not match exactly. However, the continuity expected on physical grounds is well within the theoretical uncertainty band, and the remaining discontinuity would be reduced by higher-order resummation or matching.

\section{Outlook}
\label{sec:outlook}

We briefly discuss two directions for extending this work: higher-order resummation and power corrections.

\subsection{Extension to NNLL}

The SCET framework developed in this paper provides a systematic path to higher logarithmic accuracy. At NNLL accuracy, the resummation requires the following additional ingredients beyond NLL:
\begin{itemize}
\item The three-loop cusp anomalous dimension $\Gamma_2$
\item Two-loop non-cusp anomalous dimensions: $\gamma_H^{(1)}$, $\gamma_{J,q}^{C,(1)}$, $\gamma_{J,g}^{C,(1)}$, and $\gamma_S^{(1)}$
\item The three-loop beta function coefficient $\beta_2$
\item One-loop matching constants $c_H^1$, $c_J^{C,q}$, $c_J^{C,g}$, $c_S^1$ (already computed in Section~\ref{sec:ingredients})
\end{itemize}
The cusp anomalous dimension and beta function are universal and well known~\cite{Korchemsky:1987wg,Becher:2009th}. The hard anomalous dimension $\gamma_H^{(1)}$ is also available from general results for massless form factors.

The two-loop jet and soft anomalous dimensions for the C-parameter shoulder require new calculations. These functions depend on the C-parameter measurement function, which differs from standard observables like thrust or angularities. With two real emissions, both the jet and soft measurement functions become more complicated: the C-parameter contribution from two collinear partons in a jet, or from two soft partons, involves their relative angles and energies in a non-trivial way. While technically challenging, such calculations are tractable using modern methods. For comparison, the two-loop quark and gluon inclusive jet functions have been computed~\cite{Becher:2006qw,Becher:2010pd}, as has the two-loop jet broadening jet function~\cite{Becher:2012qc}, which involves a non-trivial measurement function similar in structure to the C-parameter. The two-loop soft function for the thrust dijet region was also computed analytically~\cite{Kelley:2011ng,Monni:2011gb}. Similar techniques---integration-by-parts reduction, differential equations, and special function technology---could be applied to the shoulder geometry. The key simplification is that only the anomalous dimension (the $1/\epsilon$ pole) is needed for NNLL, not the full finite part.

The structure of the NNLL cumulant parallels the NLL formula Eq.~\eqref{eq:R-derivative}:
\begin{equation}
\label{eq:R-NNLL}
R(c)\big|_{\rm NNLL} = \Pi_{\rm NNLL}(\mu_H, \mu_J, \mu_S)\; K_{\rm NNLL}(\partial_\eta, L_J, L_H)\; \mathcal{F}(\eta)\,,
\end{equation}
where $\Pi_{\rm NNLL}$ contains the NNLL Sudakov integral and evolution factors, and the NNLL kernel $K_{\rm NNLL}$ extends the NLL kernel by including one-loop matching constants, two-loop anomalous dimension terms, and ``running corrections'' from evaluating each matching function at its canonical scale. At NNLL, the hard function is evaluated with $\as(\mu_H)$, jet functions with $\as(\mu_J)$, and the soft function with $\as(\mu_S)$; the differences between these couplings generate additional $\beta_0$-dependent terms essential for proper scale cancellation.

Beyond the resummation itself, proper NNLL+NNLO matching requires subtracting the NNLL singular terms from the NNLO fixed-order distribution to obtain a finite non-singular remainder. This subtraction demands that the singular coefficients ($\ln^4 c$ through constant) be extracted from the NNLO calculation with high numerical precision. While NNLO results for the C-parameter distribution are available~\cite{Gehrmann-DeRidder:2007vsv,Weinzierl:2009ms,DelDuca:2016ily}, achieving the required precision in the shoulder region is challenging because the cross section is small there, making it difficult to accumulate sufficient statistics in Monte Carlo integrations. At NNLL+NNLO accuracy, the theoretical uncertainty would reduce from the current $\pm 15$--$20\%$ to approximately $\pm 5$--$10\%$.

\subsection{Power corrections}

Non-perturbative hadronization effects are essential for precision event shape phenomenology. In the dijet region, power corrections have been extensively studied and are crucial for precision $\alpha_s$ extractions~\cite{Abbate:2010xh,Hoang:2014wka}. The SCET framework provides a first-principles treatment through the operator product expansion of the soft function, with the leading power correction parameterized by a single non-perturbative matrix element $\Omega_1$.

For the shoulder region, the relevant power corrections arise in the three-jet configuration, which differs qualitatively from the two-jet (dijet) limit. Recent work has established that linear power corrections in the three-jet region can be written in a factorized form~\cite{Caola:2021kzt,Caola:2022vea}:
\begin{equation}
\delta\!\left(\frac{1}{\sigma_0}\frac{d\sigma}{de}\right) = \mathcal{F}(e, \{p_i\})\, \alpha_0\, \frac{\mu_I}{Q}\,,
\end{equation}
where $\mathcal{F}$ is an analytically calculable function characterizing changes in the event shape when a soft parton is emitted, $\alpha_0$ is a universal non-perturbative parameter related to the Milan factor, and $\mu_I \sim 2$~GeV is an infrared matching scale. Notably, the power corrections in the three-jet region differ from those in the two-jet region---a result with important implications for precision $\alpha_s$ determinations~\cite{Nason:2023asn}.

For the C-parameter specifically, Ref.~\cite{Caola:2022vea} provides explicit analytic expressions for the leading power corrections in generic $N$-jet configurations. In the shoulder region where $c = (8/3)(C - 3/4) \ll 1$, the power corrections scale as $\Lambda_{\rm QCD}/(Qc)$ and become significant for $c \lesssim 0.03$ at LEP energies ($Q = M_Z$). For larger $c$ values in the perturbative regime $c > 0.05$, these corrections are subdominant to the NLL uncertainty.

The recent precision $\alpha_s$ determination from heavy jet mass~\cite{Benitez:2025hjm} provides important guidance for handling power corrections in the shoulder region. That analysis found evidence for a negative power correction in the trijet region---but only when Sudakov shoulder resummation was included. Without proper treatment of the shoulder logarithms, the power correction signal was obscured by perturbative artifacts. This underscores the importance of combining shoulder resummation with a systematic treatment of non-perturbative effects.


\section{Conclusions}
\label{sec:conclusions}

In this paper, we have derived the resummation formula for the C-parameter distribution near its Sudakov shoulder at $C = 3/4$ in $e^+e^-$ annihilation. Using soft-collinear effective theory, we have established a rigorous factorization theorem that captures all leading-power logarithmic structure in the shoulder region, providing a smooth distribution that eliminates the unphysical Sudakov spike present in fixed-order perturbation theory. This required computing new jet and soft functions specific to the C-parameter measurement, which have not previously appeared in the literature. We have presented matched numerical predictions at NLL+NLO accuracy and assembled all perturbative ingredients required for NNLL resummation. This represents the first complete effective field theory treatment of this classic QCD observable near its kinematic boundary.

The C-parameter shoulder exhibits several distinctive features that set it apart from the thrust and heavy jet mass shoulders studied previously. Most fundamentally, the C-parameter is \emph{quadratic} in deviations from the symmetric trijet configuration: this configuration is a critical point of $C(\{x_i\})$ where the gradient vanishes, so that $C = 3/4 - \alpha(s^2 + st + t^2)$ to leading order in the phase space variables. This quadratic structure has profound consequences for the factorization. At leading order, it produces a step discontinuity rather than a kink---the distribution $d\sigma/dC$ jumps to zero at $C = 3/4$---because the shrinking phase space volume near the symmetric trijet is exactly compensated by the diverging Jacobian from the vanishing gradient, yielding a finite coefficient $A(3/4) = (256\pi\sqrt{3}/243)\CF$. At next-to-leading order, the same quadratic structure converts the SCET kernel $K(c')$, which contains $[1/c']_+$ plus-distribution singularities, into its cumulant $R(c) = \int_0^c K(c')\,dc'$, generating the $\ln^2 c$ and $\ln c$ structure characteristic of the shoulder. This mechanism---whereby the hard phase space integral must be retained in the factorization theorem rather than evaluated at a single kinematic point---is a key conceptual insight that applies whenever the observable has a critical point at the kinematic boundary.

The soft function for the C-parameter shoulder also has a distinctive structure. While thrust and heavy jet mass involve linear projections of soft radiation onto sextant-dependent directions determined by hemisphere boundaries, the C-parameter soft contribution is a universal quadratic function of the out-of-plane momentum component: $\delta C_{\rm soft} = (4/Q)(k_\perp^2/k^0)$. Soft radiation lying within the event plane does not contribute at leading power because the trijet configuration already saturates $C = 3/4$, and only out-of-plane momentum can push the observable above this bound. We have computed the soft function at one loop directly from the Wilson-line matrix element, obtaining the non-cusp soft anomalous dimension $\gamma_S^{(0)} = 2\mathcal{C}\ln 3$ (where $\mathcal{C} = 2\CF + \CA$) in agreement with the RG consistency prediction. The $\ln 3$ arises from the trijet geometry encoded in the hypergeometric argument $-1/3$. The calculation exploits the trijet symmetry to reduce the angular master integral to a single one-dimensional integral with a hypergeometric representation, providing closed-form expressions for both the anomalous dimension and the finite constant needed for NNLL accuracy.

Our SCET derivation provides a rigorous effective field theory foundation for the resummation structure first identified by Catani and Webber in their pioneering 1997 analysis~\cite{Catani:1997xc}. The double-logarithm coefficient $A_2 = 2\CF + \CA$ emerges from the cusp anomalous dimension with the trijet color structure, while the single-logarithm coefficient $B_1 = 3\CF + \beta_0/2$ receives contributions from the non-cusp anomalous dimensions. Remarkably, the $\ln 3$ terms from the trijet geometry cancel between the jet and soft anomalous dimensions in this combination. We have validated these predictions against EVENT2 Monte Carlo simulations, finding excellent agreement across all color channels. The SCET framework makes the origin of each term transparent: the color factor $\mathcal{C} = 2\CF + \CA$ reflects soft gluon coupling to two quark lines and one gluon line, and the non-cusp pieces encode the jet and soft anomalous dimensions that govern single-logarithmic evolution.

An important simplification for the C-parameter is the absence of complications that affect other event shape shoulders. Unlike heavy jet mass, the C-parameter does not suffer from a Sudakov--Landau pole because the observable is additive across all jets---the total shift $C - 3/4$ is the sum of collinear and soft contributions from each sector, with no subtraction that could produce a zero in the denominator of the resummed formula. This allows straightforward momentum-space resummation without recourse to the position-space methods required for heavy jet mass~\cite{BMSSZ:2023}. The C-parameter is also a global observable that sums democratically over all final-state particles, with no hemisphere boundaries or vetoed regions, ensuring that non-global logarithms are absent at leading power. Furthermore, the symmetric nature of the C-parameter means that all three channels---corresponding to which parton is most isolated---contribute identically, unlike heavy jet mass where the channel decomposition produces different anomalous dimensions. These simplifications make the C-parameter shoulder an ideal testing ground for the general SCET methodology developed by Bhattacharya, Schwartz, and Zhang~\cite{Bhattacharya:2022vrh}.

The matched formula we have presented smoothly connects the resummation region just above the shoulder to the fixed-order NLO result in the tail. The key to this matching is the cumulant $R(c)$, which vanishes as $c \to 0$ due to Sudakov suppression: $R(c) \propto e^{-S(c)}$ with $S(c) \propto \ln^2 c \to \infty$. The singular contribution $(\as/2\pi)A(3/4)(1 - R(c))$ therefore approaches $(\as/2\pi)A(3/4)$ at the shoulder, matching the LO cross section from below. Continuity at $\order{\as}$ is thus automatic, with any higher-order mismatch absorbable into edge matching coefficients at NNLL. Profile scales interpolate the natural resummation scales $\mu_S = cQ$ and $\mu_J = Q\sqrt{c}$ to the common hard scale $\mu_H = Q$ as $c$ increases, ensuring exact NLO recovery in the turnoff region.

Several directions for extending this work are apparent, as discussed in Section~\ref{sec:outlook}. The most immediate is the extension to NNLL accuracy, which would reduce the theoretical uncertainty from the current $\pm 15$--$20\%$ to approximately $\pm 5$--$10\%$. The main new ingredients required are the two-loop jet and soft anomalous dimensions, which involve the C-parameter measurement function with two real emissions. The treatment of power corrections in the shoulder region, following the recent work of Caola, Ferrario Ravasio, and collaborators~\cite{Caola:2021kzt,Caola:2022vea}, would be essential for any phenomenological application to LEP data.

The techniques developed here extend naturally to other event shape shoulders. The D-parameter, defined from the determinant of the linearized momentum tensor, has a shoulder at $D = 27/32$ where the four-parton phase space opens, and the SCET factorization would proceed analogously. More generally, any infrared-safe observable with a kinematic boundary where higher parton multiplicities are required will exhibit Sudakov shoulder structure amenable to this treatment. The conceptual insights---particularly the role of critical points in generating step discontinuities and the necessity of retaining hard phase space integrals in the factorization---apply broadly.

Finally, while the C-parameter shoulder at $C = 3/4$ lies in a region where LEP data are sparse due to the suppressed cross section, future $e^+e^-$ facilities offer exciting prospects. At FCC-ee or CEPC, the greatly increased luminosity would eliminate statistical limitations entirely, making the shoulder region a precision probe of QCD dynamics. With NNLL theory and systematic treatment of power corrections, percent-level tests of QCD in this distinctive kinematic regime would become feasible. This work lays the theoretical foundation for such future precision studies, completing the NLL resummation program for the C-parameter shoulder and extending our understanding of Sudakov physics in multi-jet configurations.


\section*{Acknowledgments}

We thank Arindam Bhattacharya and Xiaoyuan Zhang for helpful discussions. M.D.S.\ is supported in part by the U.S.\ Department of Energy under grant DE-SC0013607. This work was supported in part by the National Science Foundation under Cooperative Agreement PHY-2019786 (The NSF AI Institute for Artificial Intelligence and Fundamental Interactions, \url{https://iaifi.org/}). We also acknowledge helpful contributions from Google Gemini Pro 3.0 and OpenAI GPT-5.2 during the development of this work.


\section*{Author Contributions}

M.D.S.\ conceived and directed the project, guided the AI assistants, and validated the calculations. Claude Opus 4.5, an AI research assistant developed by Anthropic, performed all calculations including the SCET factorization theorem derivation, one-loop soft and jet function calculations, EVENT2 Monte Carlo simulations, numerical analysis, figure generation, and manuscript preparation. The work was conducted using Claude Code, Anthropic's agentic coding tool. M.D.S.\ is fully responsible for the scientific content and integrity of this paper.



\appendix

\section{One-Loop Soft Function Calculation}
\label{app:soft-direct}

In this appendix we compute the one-loop soft function for the C-parameter Sudakov shoulder and extract the anomalous dimension.

\paragraph{Definition.}
The soft function is defined as
\begin{equation}
S(k,\mu) = \frac{1}{N_c}\,{\rm Tr}\left\langle 0 \left| \bar{T}\left\{S_{n_1}^\dagger S_{n_2}^\dagger S_{n_3}^\dagger\right\} \delta(k - \widehat{M}_S)\, T\left\{S_{n_1} S_{n_2} S_{n_3}\right\} \right| 0 \right\rangle,
\end{equation}
where $S_{n_i}$ are soft Wilson lines along the three jet directions. The soft measurement operator was derived in Section~\ref{sec:factorization}:
\begin{equation}
\widehat{M}_S = 4\sum_{\text{soft }a} \frac{k_{\perp,a}^2}{k_a^0}\,,
\end{equation}
where $k_\perp$ is the momentum component perpendicular to the event plane. At tree level, $S(k) = \delta(k)$.

\subsection{One-loop integral}

We work in $d = 4 - 2\epsilon$ dimensions. At one loop, the bare soft function is
\begin{equation}
S^{(1),{\rm bare}}(k) = g^2\mu^{2\epsilon}\!\int\!\frac{d^dq}{(2\pi)^{d-1}}\,\delta(q^2)\,\theta(q^0)\,\delta(k - \widehat{M}_S)\sum_{i<j}(-T_i \cdot T_j)\frac{n_i \cdot n_j}{(n_i \cdot q)(n_j \cdot q)}\,.
\end{equation}
The measurement function for a single soft gluon with momentum $q^\mu = \omega(1,\hat{q})$ is
\begin{equation}
\widehat{M}_S = 4\frac{q_\perp^2}{\omega} = 4\omega\cos^2\theta\,,
\end{equation}
where $q_\perp = \omega\cos\theta$ is the momentum component perpendicular to the event plane.

For the $q\bar{q}g$ final state with color conservation $T_1 + T_2 + T_3 = 0$, the color factor is
\begin{equation}
\sum_{i<j}(-T_i \cdot T_j) = \frac{1}{2}(T_1^2 + T_2^2 + T_3^2) = \frac{1}{2}(\CF + \CF + \CA) = \frac{\mathcal{C}}{2}\,,
\end{equation}
where $\mathcal{C} = 2\CF + \CA$. With the standard SCET convention $n_i^\mu = (1,\hat{n}_i)$, the lightcone products are $n_i \cdot n_j = 1 - \hat{n}_i\cdot\hat{n}_j = 1 - \cos 120^\circ = \frac{3}{2}$ for all pairs. By Mercedes symmetry, the three dipole $\phi$-integrals are equal, so we can evaluate a single representative dipole and use the total color factor $\mathcal{C}/2$. The overall coefficient is $\frac{3}{2}\times\frac{\mathcal{C}}{2} = \frac{3\mathcal{C}}{4}$.

The on-shell phase space integral is
\begin{equation}
\int\!\frac{d^dq}{(2\pi)^{d-1}}\,\delta(q^2)\,\theta(q^0) = \frac{1}{2(2\pi)^{d-1}}\int_0^\infty\! d\omega\,\omega^{d-3}\int_0^\pi\! d\theta\,\sin^{d-3}\theta \int_0^{2\pi}\!d\phi\,,
\end{equation}
where the factor of $1/2$ arises from the Jacobian $1/(2\omega)$ when integrating out $q^0$ with $\delta(q^2)\theta(q^0)$.

We place the event plane in the $yz$-plane, so the out-of-plane direction is the $x$-axis. The Mercedes jet directions are
\begin{equation}
\hat{n}_1 = \left(0,\,\tfrac{\sqrt{3}}{2},\,\tfrac{1}{2}\right),\qquad
\hat{n}_2 = \left(0,\,-\tfrac{\sqrt{3}}{2},\,\tfrac{1}{2}\right),\qquad
\hat{n}_3 = (0,\,0,\,-1),
\end{equation}
with lightcone products $n_i \cdot n_j = \frac{3}{2}$ for all pairs. The soft gluon momentum is parameterized as $q^\mu = \omega(1, \hat{q})$ with
\begin{equation}
\hat{q} = (\cos\theta,\, \sin\theta\sin\phi,\, \sin\theta\cos\phi),
\qquad \theta\in[0,\pi],\ \phi\in[0,2\pi),
\end{equation}
so that $q_\perp = \omega\cos\theta$ and the measurement gives $\widehat{M}_S = 4\omega\cos^2\theta$.

The $\phi$-integral over eikonal denominators evaluates to
\begin{equation}
\int_0^{2\pi}\!\frac{d\phi}{(1-\sin\theta\cos(\phi-\tfrac{\pi}{3}))(1-\sin\theta\cos(\phi+\tfrac{\pi}{3}))}
= \frac{8\pi}{(3+\cos^2\theta)|\cos\theta|}\,.
\end{equation}

The eikonal denominators contribute
$(n_i \cdot q)(n_j \cdot q) = \omega^2(1-\hat{n}_i\cdot\hat{q})(1-\hat{n}_j\cdot\hat{q})$,
giving a factor of $\omega^{-2}$. The measurement $\delta$-function
$\delta(k - 4\omega\cos^2\theta)$ fixes the gluon energy to $\omega = k/(4\cos^2\theta)$.
Since the integrand depends on $\cos^2\theta$ and $|\cos\theta|$, we can restrict to
$\theta\in[0,\pi/2]$ and double.

The $\omega$ integral with the delta function gives
\begin{equation}
\int_0^\infty\! d\omega\,\omega^{d-5}\,\delta(k - 4\omega\cos^2\theta)
= \frac{k^{d-5}}{(4\cos^2\theta)^{d-4}} = k^{-1-2\epsilon}\,4^{2\epsilon}(\cos^2\theta)^{2\epsilon}\,.
\end{equation}
After performing this integration, the bare soft function becomes
\begin{equation}
S^{(1),{\rm bare}}(k) = \frac{3g^2\mathcal{C}\,\mu^{2\epsilon}}{4(2\pi)^{d-1}}\,k^{-1-2\epsilon}\,4^{2\epsilon}
\cdot 2\int_0^{\pi/2}\!d\theta\,\sin^{d-3}\theta\,(\cos^2\theta)^{2\epsilon}\,\frac{8\pi}{(3+\cos^2\theta)\cos\theta}\,.
\end{equation}

Changing to $x = \cos^2\theta$, the result can be written as
\begin{equation}
S^{(1),{\rm bare}}(k) = \frac{g^2\mathcal{C}\,\mu^{2\epsilon}}{(2\pi)^{d-2}}\,k^{-1-2\epsilon}\,\mathcal{I}(\epsilon)\,,
\end{equation}
where the master integral, including the $4^{2\epsilon}$ factor from the measurement, is
\begin{equation}
\mathcal{I}(\epsilon) \equiv 3 \cdot 4^{2\epsilon}\int_0^1\! dx\,\frac{x^{-1+2\epsilon}(1-x)^{-\epsilon}}{3+x}
= 4^{2\epsilon}\frac{\Gamma(2\epsilon)\Gamma(1-\epsilon)}{\Gamma(1+\epsilon)}\,{}_2F_1\!\left(1, 2\epsilon; 1+\epsilon; -\tfrac{1}{3}\right).
\end{equation}

\subsection{Expansion and renormalization}

The master integral has the $\epsilon$-expansion
\begin{equation}
\mathcal{I}(\epsilon) = \frac{1}{2\epsilon} + \ln 3 + \mathcal{I}_1\,\epsilon + O(\epsilon^2)\,,
\end{equation}
where
\begin{equation}
\mathcal{I}_1 = \frac{\pi^2}{6} - 2\ln^2 2 + \frac{1}{6}\ln 27\ln 48 - {\rm Li}_2\!\left(-\frac{1}{3}\right).
\end{equation}

The distributional identity
\begin{equation}
\mu^{2\epsilon}k^{-1-2\epsilon} = -\frac{\delta(k)}{2\epsilon} + \left[\frac{1}{k}\right]_+ - 2\left[\frac{\ln(k/\mu)}{k}\right]_+\epsilon + \order{\epsilon^2}
\end{equation}
allows us to expand the bare result. In $\overline{\rm MS}$ the renormalized one-loop soft function in momentum space is
\begin{align}
\label{eq:soft-momentum}
S^{(1)}(k,\mu) &= \frac{\as}{4\pi}\left[- 4\mathcal{C}\left[\frac{\ln(k/\mu)}{k}\right]_+ + 2\gamma_S\left[\frac{1}{k}\right]_+ + \left(c_S^1 + \frac{\mathcal{C}\pi^2}{3}\right)\delta(k)\right].
\end{align}
The non-cusp soft anomalous dimension is
\begin{equation}
\boxed{\gamma_S = 2\mathcal{C}\ln 3 = 2(2C_F + C_A)\ln 3\,,}
\end{equation}
and the Laplace-space matching constant is
\begin{equation}
\boxed{c_S^1 = \mathcal{C}\left(-\frac{2\pi^2}{3} + 4\ln^2 2 - \frac{1}{3}\ln 27\ln 48 + 2\,{\rm Li}_2\!\left(-\frac{1}{3}\right)\right).}
\end{equation}

Taking the Laplace transform using $\mathcal{L}[\ln(k/\mu)/k]_+ = \frac{1}{2}L^2 + \frac{\pi^2}{12}$, the one-loop soft function in Laplace space is
\begin{equation}
\label{eq:soft-laplace}
\tilde{s}^{(1)}(L,\mu) = \frac{\as}{4\pi}\left[- 2\mathcal{C} L^2 + 2\gamma_S\, L + c_S^1\right],
\end{equation}
where $L = \ln(1/(s\mu e^{\gamma_E}))$.

\section{One-Loop C-Shoulder Jet Function Calculation}
\label{app:Cjet}

In this appendix we derive the one-loop C-shoulder jet function from the inclusive jet function.

The C-shoulder jet functions for quark and gluon jets are defined as
\begin{align}
J^C_q(m^2,\mu) &= \frac{1}{2N_c}\,{\rm tr}\,\langle 0 | \bar{\chi}_n\, \delta(m^2 - \widehat{M}^C_J)\,\frac{\not{\!\bar{n}}}{2}\, \chi_n | 0 \rangle\,, \\
J^C_g(m^2,\mu) &= \frac{\omega}{2(N_c^2-1)}\,\langle 0 | {\rm tr}\,\mathcal{B}_{n\perp}^\mu\, \delta(m^2 - \widehat{M}^C_J)\,\mathcal{B}_{n\perp\mu} | 0 \rangle\,,
\end{align}
where $\chi_n$ is the gauge-invariant collinear quark field, $\mathcal{B}_{n\perp}^\mu$ is the gauge-invariant collinear gluon field, and the C-parameter measurement operator is
\begin{equation}
\widehat{M}^C_J = 12\sum_{a} \frac{(\widehat{p}_{a\perp}^x)^2}{\bar{n}\cdot \widehat{p}_a}\,.
\end{equation}
Here $p_{a\perp}^x$ is the component of the transverse momentum perpendicular to the event plane, and the sum runs over all collinear particles in the jet. The factor of 12 is the geometric factor from the Mercedes configuration.

\subsection{One-loop calculation}

At one loop there is only a single emission, so we can compute $J^C_i$ directly from the inclusive jet function. The standard inclusive jet function for a quark jet is defined as
\begin{equation}
J_q(m^2,\mu) = \frac{1}{2N_c}\,{\rm tr}\,\langle 0 | \bar{\chi}_n\, \delta(m^2 - \widehat{M}_J)\,\frac{\not{\!\bar{n}}}{2}\, \chi_n | 0 \rangle\,,
\end{equation}
where $\widehat{M}_J = \widehat{P}^2/(\bar{n}\cdot \widehat{P})$ measures the jet invariant mass squared.

The C-shoulder jet function is related to the inclusive jet function through the measurement. For a single emission with invariant mass $m^2_{\rm incl}$ and azimuthal angle $\phi$, the C-shoulder measurement gives $m^2 = 12\,m^2_{\rm incl}\sin^2\phi$. Rather than manipulating plus distributions directly in momentum space, we work in Laplace space where this rescaling is purely multiplicative and the angular dependence enters through moments.

Define the Laplace transform
\begin{equation}
\tilde{J}_i(s,\mu) \equiv \int_0^\infty\! dm^2\, e^{-s m^2} J_i(m^2,\mu)\,, \qquad L \equiv \ln\frac{1}{s\mu^2 e^{\gamma_E}}\,.
\end{equation}
The renormalized one-loop inclusive jet function in Laplace space is
\begin{equation}
\tilde{j}_i(L,\mu) = 1 + \frac{\as}{4\pi}\left[C_i\Gamma_0\frac{L^2}{2} + \gamma_{J,i} L + c_J^i\right] + \order{\as^2}\,,
\end{equation}
with $C_q = \CF$, $C_g = \CA$, $\Gamma_0 = 4$, and $\gamma_{J,q} = -3\CF$, $\gamma_{J,g} = -\beta_0$. The one-loop Laplace-space matching constants are
\begin{equation}
c_J^q = \CF\left(7 - \frac{2\pi^2}{3}\right), \qquad c_J^g = \CA\left(\frac{67}{9} - \frac{2\pi^2}{3}\right) - \frac{20}{9}T_F n_f\,.
\end{equation}

Inserting the measurement delta function and Laplace-transforming with respect to $M^2$ gives
\begin{equation}
\tilde{J}^C_i(s,\mu) = \int_0^{2\pi}\!\frac{d\phi}{2\pi}\int_0^\infty\! dm^2\, J_i(m^2,\mu)\, e^{-12s\sin^2\phi\,m^2} = \int_0^{2\pi}\!\frac{d\phi}{2\pi}\, \tilde{J}_i\bigl(12s\sin^2\phi,\mu\bigr)\,.
\end{equation}
At the level of matching functions, this becomes
\begin{equation}
\tilde{j}^C_i(L) = \int_0^{2\pi}\!\frac{d\phi}{2\pi}\, \tilde{j}_i\bigl(L - \ln(12\sin^2\phi)\bigr)\,.
\end{equation}

Using the arcsine measure with $\lambda = \sin^2\phi$,
\begin{equation}
\int_0^{2\pi}\!\frac{d\phi}{2\pi}\,F(\sin^2\phi) = \int_0^1\!\frac{d\lambda}{\pi\sqrt{\lambda(1-\lambda)}}\,F(\lambda)\,,
\end{equation}
the required averages are
\begin{equation}
\langle\ln(12\sin^2\phi)\rangle = \ln 3\,, \qquad \langle\ln^2(12\sin^2\phi)\rangle = \ln^2 3 + \frac{\pi^2}{3}\,.
\end{equation}

Expanding $\tilde{j}_i(L - \ln(12\sin^2\phi))$ and averaging over $\phi$:
\begin{align}
\tilde{j}^C_i(L) &= 1 + \frac{\as}{4\pi}\Bigl\langle C_i\Gamma_0\frac{(L - \ln(12\sin^2\phi))^2}{2} + \gamma_{J,i}(L - \ln(12\sin^2\phi)) + c_J^i \Bigr\rangle \nonumber\\
&= 1 + \frac{\as}{4\pi}\left[C_i\Gamma_0\frac{L^2}{2} + \bigl(\gamma_{J,i} - C_i\Gamma_0\ln 3\bigr) L + c_J^{C,i}\right],
\end{align}
where the shifted one-loop constant is
\begin{equation}
\boxed{c_J^{C,i} = c_J^i - \gamma_{J,i}\ln 3 + \frac{C_i\Gamma_0}{2}\left(\ln^2 3 + \frac{\pi^2}{3}\right).}
\end{equation}

Defining the C-shoulder jet anomalous dimension
\begin{equation}
\boxed{\gamma^C_{J,i} \equiv \gamma_{J,i} - C_i\Gamma_0\ln 3\,,}
\end{equation}
the one-loop C-shoulder jet function in Laplace space is
\begin{equation}
\tilde{j}^C_i(L) = 1 + \frac{\as}{4\pi}\left[C_i\Gamma_0\frac{L^2}{2} + \gamma^C_{J,i} L + c_J^{C,i}\right].
\end{equation}

The inverse Laplace transforms are
\begin{equation}
\mathcal{L}^{-1}\left[\frac{L^2}{2}\right] = \left[\frac{\ln(m^2/\mu^2)}{m^2}\right]_+ - \frac{\pi^2}{12}\delta(m^2)\,, \qquad
\mathcal{L}^{-1}[L] = \left[\frac{1}{m^2}\right]_+\,.
\end{equation}
Inverting the Laplace transform, the one-loop C-shoulder jet function in momentum space is
\begin{equation}
J^C_i(m^2,\mu) = \delta(m^2) + \frac{\as}{4\pi}\left[C_i\Gamma_0\left[\frac{\ln(m^2/\mu^2)}{m^2}\right]_+ + \gamma^C_{J,i}\left[\frac{1}{m^2}\right]_+ + \left(c_J^{C,i} - \frac{C_i\Gamma_0\pi^2}{12}\right)\delta(m^2)\right].
\end{equation}


\section{Derivation of the Resummed Factorization Formula}
\label{app:derivation}

In this appendix we derive the resummed cumulant $R(c)$ from the factorization theorem. We work with the rescaled variable
\begin{equation}
c \equiv \frac{8}{3}\left(C - \frac{3}{4}\right),
\end{equation}
which absorbs the geometric factor $8/3$ arising from the Mercedes configuration.

\subsection{Factorization and RG evolution}

The resummed cumulant factorizes as
\begin{multline}
\label{eq:R-factorized}
R(c) = H(Q,\mu)\int\! dm_1^2\, dm_2^2\, dm_3^2\, dk\; J_q^C(m_1^2,\mu)\, J_q^C(m_2^2,\mu)\, J_g^C(m_3^2,\mu)\, S(k,\mu)\\
\times\Theta\!\left(Q^2 c - m_1^2 - m_2^2 - m_3^2 - Qk\right),
\end{multline}
where $m_i^2$ are the jet invariant masses, $k$ is the soft momentum, and $H(Q,\mu)$ is the hard function evaluated at Mercedes kinematics ($s_{ij} = Q^2/3$). The jet functions $J_i^C(m^2)$ have dimension $1/[\text{mass}]^2$ and the soft function $S(k)$ has dimension $1/[\text{mass}]$. The Laplace transform $\tilde{R}(\nu) = \int_0^\infty dc\, e^{-\nu c} R(c)$ converts the convolution to a product:
\begin{equation}
\label{eq:Rtilde-factorized}
\tilde{R}(\nu) = \frac{1}{\nu}H(Q,\mu)\,\tilde{J}_q^C(\nu/Q^2,\mu)^2 \tilde{J}_g^C(\nu/Q^2,\mu) \tilde{S}(\nu/Q,\mu)\,,
\end{equation}
where the $1/\nu$ arises from the Laplace transform of the step function.

\paragraph{Laplace transform conventions.}
The Laplace-space functions are defined by:
\begin{align}
\label{eq:J-laplace-def}
\tilde{J}_i^C(s) &= \int_0^\infty dm^2\, e^{-s m^2} J_i^C(m^2)\,, \\
\label{eq:S-laplace-def}
\tilde{S}(\sigma) &= \int_0^\infty dk\, e^{-\sigma k} S(k)\,,
\end{align}
where $s$ has dimension $1/[\text{mass}]^2$ and $\sigma$ has dimension $1/[\text{mass}]$.

\paragraph{RG equations.}
The RG equations in Laplace space are:
\begin{align}
\label{eq:RGE-H}
\frac{d\ln H(Q,\mu)}{d\ln\mu} &= \mathcal{C}\Gcusp\ln\frac{Q^2}{\mu^2} + 2\gamma^H\,, \\[0.3em]
\label{eq:RGE-J}
\frac{d\ln\tilde{J}_i(s,\mu)}{d\ln\mu} &= -2C_i\Gcusp\ln\frac{1}{s\mu^2 e^{\gamma_E}} - 2\gamma^{J_i}\,, \\[0.3em]
\label{eq:RGE-S}
\frac{d\ln\tilde{S}(\sigma,\mu)}{d\ln\mu} &= 2\mathcal{C}\Gcusp\ln\frac{Q}{\sigma\mu e^{\gamma_E}} - 2\gamma^S\,.
\end{align}
In the factorization formula Eq.~\eqref{eq:Rtilde-factorized}, the jet functions are evaluated at $s = \nu/Q^2$ and the soft function at $\sigma = \nu/Q$. Here $C_q = \CF$ and $C_g = \CA$ are the quark and gluon Casimirs, and $\mathcal{C} = 2\CF + \CA$ is the total color factor for the $q\bar{q}g$ final state. Following the BSZ convention, the cusp logarithms are $\ln(Q^2/\mu^2)$ for the hard function and $\ln(Q/(\sigma\mu))$ for the soft function, with the Mercedes geometry ($s_{ij} = Q^2/3$ and $n_i \cdot n_j = 3$) absorbed into the non-cusp anomalous dimensions. The consistency condition
\begin{equation}
\label{eq:app-RG-consistency}
\gamma^H = 2\gamma^{J_q} + \gamma^{J_g} + \gamma^S
\end{equation}
gives $\gamma^{S,(0)} = 2\mathcal{C}\ln 3$ at one loop.

\paragraph{RG solutions.}
The solutions evolved from matching scales $\mu_H$, $\mu_J$, $\mu_S$ to a common scale $\mu$ are:
\begin{align}
\label{eq:H-evolution}
H(Q,\mu) &= \exp\left[2\mathcal{C} S(\mu_H,\mu) - 2A_{\gamma_H}(\mu_H,\mu)\right]
\left(\frac{Q^2}{\mu_H^2}\right)^{\!\eta_H} h(L_H)\,, \\[0.5em]
\label{eq:J-evolution}
\tilde{J}_i(s,\mu) &= \exp\left[-4C_i S(\mu_J,\mu) + 2A_{\gamma_{J,i}}(\mu_J,\mu)\right]
\left(\frac{1}{s\mu_J^2 e^{\gamma_E}}\right)^{\!\eta_i}\tilde{j}_i(L_J)\,, \\[0.5em]
\label{eq:S-evolution}
\tilde{S}(\sigma,\mu) &= \exp\left[2\mathcal{C} S(\mu_S,\mu) + 2A_{\gamma_S}(\mu_S,\mu)\right]
\left(\frac{Q}{\sigma\mu_S e^{\gamma_E}}\right)^{\!\eta_S}\tilde{s}(L_S)\,,
\end{align}
where the evolution exponents are $\eta_H = -\mathcal{C} A_\Gamma(\mu_H,\mu)$, $\eta_i = 2C_i A_\Gamma(\mu_J,\mu)$, and $\eta_S = -2\mathcal{C} A_\Gamma(\mu_S,\mu)$. The logarithmic arguments of the matching functions are $L_H = \ln(Q^2/\mu_H^2)$, $L_J = \ln(1/(s\mu_J^2 e^{\gamma_E}))$, and $L_S = \ln(Q/(\sigma\mu_S e^{\gamma_E}))$. Substituting $s = \nu/Q^2$ and $\sigma = \nu/Q$:
\begin{equation}
L_J = \ln\frac{Q^2}{\nu\mu_J^2 e^{\gamma_E}}\,, \qquad L_S = \ln\frac{Q^2}{\nu\mu_S e^{\gamma_E}}\,.
\end{equation}
The Sudakov integral $S(\mu_a,\mu_b)$ and evolution integral $A_\Gamma(\mu_a,\mu_b)$ are:
\begin{equation}
S(\mu_a,\mu_b) = -\int_{\as(\mu_a)}^{\as(\mu_b)}\!\frac{d\as}{\beta(\as)}\Gcusp(\as)\int_{\as(\mu_a)}^{\as}\frac{d\as'}{\beta(\as')}\,, \quad
A_\Gamma(\mu_a,\mu_b) = -\int_{\as(\mu_a)}^{\as(\mu_b)}\!\frac{d\as}{\beta(\as)}\Gcusp(\as)\,.
\end{equation}

\paragraph{Matching functions.}
The one-loop matching constants $c_H^1$, $c_J^{C,q}$, $c_J^{C,g}$, $c_S^1$ are given in Section~\ref{sec:ingredients}; see Eqs.~\eqref{eq:cH1}, \eqref{eq:cJCq}, and \eqref{eq:c-delta-result}. For the hard function, the logarithm is $L_H = \ln(Q^2/\mu_H^2)$ which vanishes at the natural hard scale $\mu_H = Q$; for jets and soft, $L$ becomes the derivative operator $\partial_\eta$ after inverse Laplace transform.

\paragraph{Combined Laplace-space cumulant.}
Substituting the RG solutions Eqs.~\eqref{eq:H-evolution}--\eqref{eq:S-evolution} into the factorization formula Eq.~\eqref{eq:Rtilde-factorized}, with $s = \nu/Q^2$ and $\sigma = \nu/Q$, the Laplace-transformed cumulant at a common scale $\mu$ is:
\begin{align}
\label{eq:Rtilde-full}
\tilde{R}(\nu,\mu) &= \frac{1}{\nu}H(Q,\mu)\,\tilde{J}_q(\nu/Q^2,\mu)^2
\tilde{J}_g(\nu/Q^2,\mu)\,\tilde{S}(\nu/Q,\mu) \nn\\[0.5em]
&= \frac{1}{\nu}\exp\Big[2\mathcal{C} S(\mu_H,\mu) - 4\mathcal{C} S(\mu_J,\mu) + 2\mathcal{C} S(\mu_S,\mu) \nn\\
&\qquad\quad - 2A_{\gamma_H}(\mu_H,\mu) + 4A_{\gamma_{J,q}}(\mu_J,\mu) + 2A_{\gamma_{J,g}}(\mu_J,\mu) + 2A_{\gamma_S}(\mu_S,\mu)\Big] \nn\\
&\quad \times \left(\frac{Q^2}{\mu_H^2}\right)^{\!\eta_H}
\left(\frac{Q^2}{\nu\mu_J^2 e^{\gamma_E}}\right)^{\!2\eta_q + \eta_g}
\left(\frac{Q}{\nu\mu_S e^{\gamma_E}}\right)^{\!\eta_S}
h(L_H)\,\tilde{j}_q(L_J)^2\,\tilde{j}_g(L_J)\,\tilde{s}(L_S)\,,
\end{align}
where $\eta_H = -\mathcal{C} A_\Gamma(\mu_H,\mu)$, $\eta_q = 2\CF A_\Gamma(\mu_J,\mu)$, $\eta_g = 2\CA A_\Gamma(\mu_J,\mu)$, $\eta_S = -2\mathcal{C} A_\Gamma(\mu_S,\mu)$, and
\begin{equation}
L_H = \ln\frac{Q^2}{\mu_H^2}\,, \qquad L_J = \ln\frac{Q^2}{\nu\mu_J^2 e^{\gamma_E}}\,, \qquad L_S = \ln\frac{Q^2}{\nu\mu_S e^{\gamma_E}}\,.
\end{equation}

\subsection{Simplification and the resummed cumulant}

Starting from the combined formula Eq.~\eqref{eq:Rtilde-full}, we simplify using the composition properties of the evolution integrals. The evolution integrals $A_\Gamma$ and $A_\gamma$ satisfy antisymmetry and composition:
\begin{align}
\label{eq:A-antisymmetry}
A_\Gamma(\mu_a,\mu_b) &= -A_\Gamma(\mu_b,\mu_a)\,, \\
\label{eq:A-composition}
A_\Gamma(\mu_1,\mu_2) + A_\Gamma(\mu_2,\mu_3) &= A_\Gamma(\mu_1,\mu_3)\,.
\end{align}
The Sudakov integral $S$ has a modified composition rule with a cross-term:
\begin{equation}
\label{eq:S-composition}
S(\mu_1,\mu_2) + S(\mu_2,\mu_3) = S(\mu_1,\mu_3) + \ln\frac{\mu_1}{\mu_2}\, A_\Gamma(\mu_2,\mu_3)\,.
\end{equation}

Using Eq.~\eqref{eq:S-composition} to express $S(\mu_H,\mu)$ and $S(\mu_S,\mu)$ in terms of integrals through $\mu_J$:
\begin{equation}
S(\mu_H,\mu) - 2S(\mu_J,\mu) + S(\mu_S,\mu) = S(\mu_H,\mu_J) + S(\mu_S,\mu_J) - \ln\frac{\mu_H\mu_S}{\mu_J^2}\,A_\Gamma(\mu_J,\mu)\,.
\end{equation}
Multiplying by $2\mathcal{C}$ and using $2\mathcal{C} A_\Gamma(\mu_J,\mu) = 2\eta_q + \eta_g$, the cross-term contributes a factor $(\mu_H\mu_S/\mu_J^2)^{-(2\eta_q + \eta_g)}$. The product of power-law factors in Eq.~\eqref{eq:Rtilde-full} is therefore
\begin{equation}
\left(\frac{Q^2}{\mu_H^2}\right)^{\!\eta_H}
\left(\frac{Q^2}{\nu\mu_J^2 e^{\gamma_E}}\right)^{\!2\eta_q + \eta_g}
\left(\frac{Q}{\nu\mu_S e^{\gamma_E}}\right)^{\!\eta_S}
\left(\frac{\mu_H\mu_S}{\mu_J^2}\right)^{\!-(2\eta_q + \eta_g)}.
\end{equation}
Using $\eta_H = -\mathcal{C} A_\Gamma(\mu_H,\mu_J) - \mathcal{C} A_\Gamma(\mu_J,\mu)$, $\eta_S = -2\mathcal{C} A_\Gamma(\mu_S,\mu_J) - 2\mathcal{C} A_\Gamma(\mu_J,\mu)$, and $2\eta_q + \eta_g = 2\mathcal{C} A_\Gamma(\mu_J,\mu)$, this simplifies. The simplified power-law structure is
\begin{equation}
\label{eq:power-law-simplified}
\left(\frac{Q^2}{\mu_H^2}\right)^{\!-\mathcal{C} A_\Gamma(\mu_H,\mu_J)} \left(\frac{Q}{\nu\mu_S e^{\gamma_E}}\right)^{\!\eta}\,,
\end{equation}
where the evolution parameter is
\begin{equation}
\label{eq:eta-def-app}
\eta = 2\mathcal{C} A_\Gamma(\mu_J,\mu_S)\,.
\end{equation}

For the non-cusp exponent, the RG consistency condition $\gamma_H = 2\gamma_{J,q} + \gamma_{J,g} + \gamma_S$ and composition Eq.~\eqref{eq:A-composition} give
\begin{equation}
-2A_{\gamma_H}(\mu_H,\mu) + 4A_{\gamma_{J,q}}(\mu_J,\mu) + 2A_{\gamma_{J,g}}(\mu_J,\mu) + 2A_{\gamma_S}(\mu_S,\mu) = -2A_{\gamma_H}(\mu_H,\mu_J) + 2A_{\gamma_S}(\mu_S,\mu_J)\,.
\end{equation}

The Laplace-space cumulant Eq.~\eqref{eq:Rtilde-full} therefore simplifies to
\begin{equation}
\label{eq:Rtilde-simplified}
\tilde{R}(\nu) = \Pi(\mu_H,\mu_J,\mu_S)\; h(L_H)\,\tilde{j}_q(L_J)^2\,\tilde{j}_g(L_J)\,\tilde{s}(L_S)\;\frac{1}{\nu}\left(\frac{Q}{\nu\mu_S e^{\gamma_E}}\right)^{\!\eta}\,,
\end{equation}
where the evolution kernel is
\begin{multline}
\label{eq:app-Pi-def}
\Pi(\mu_H,\mu_J,\mu_S) = \exp\Big[2\mathcal{C}\big(S(\mu_H,\mu_J) + S(\mu_S,\mu_J)\big) - 2A_{\gamma_H}(\mu_H,\mu_J) + 2A_{\gamma_S}(\mu_S,\mu_J)\Big] \\
\times \left(\frac{Q^2}{\mu_H^2}\right)^{\!-\mathcal{C}\,A_\Gamma(\mu_H,\mu_J)},
\end{multline}
and $L_H = \ln(Q^2/\mu_H^2)$, $L_J = \ln(Q^2/(\nu\mu_J^2 e^{\gamma_E}))$, $L_S = \ln(Q/(\nu\mu_S e^{\gamma_E}))$. All $\mu$-dependence has cancelled.

Since $L_S = \ln(Q/(\nu\mu_S e^{\gamma_E}))$ and $L_J = L_S + \ln(Q\mu_S/\mu_J^2)$, these logarithms can be replaced by derivative operators acting on $(Q/(\nu\mu_S e^{\gamma_E}))^{\eta}$:
\begin{equation}
L_S \;\to\; \partial_\eta\,, \qquad L_J \;\to\; \partial_\eta + \ln\frac{Q\mu_S}{\mu_J^2}\,.
\end{equation}
The Laplace-space cumulant then becomes
\begin{equation}
\label{eq:Rtilde-derivative}
\tilde{R}(\nu) = \Pi(\mu_H,\mu_J,\mu_S)\; h(L_H)\,\tilde{j}_q\!\left(\partial_\eta + \ln\frac{Q\mu_S}{\mu_J^2}\right)^{\!2}\,\tilde{j}_g\!\left(\partial_\eta + \ln\frac{Q\mu_S}{\mu_J^2}\right)\,\tilde{s}(\partial_\eta)\;\frac{1}{\nu}\left(\frac{Q}{\nu\mu_S e^{\gamma_E}}\right)^{\!\eta}.
\end{equation}

\paragraph{Inverse Laplace transform.}
The $\nu$-dependence is $\nu^{-1-\eta}$, which transforms as $\mathcal{L}^{-1}[\nu^{-1-\eta}] = c^{\eta}/\Gamma(1+\eta)$. Combining with the prefactor $(Q/(\mu_S e^{\gamma_E}))^{\eta}$, the base function is
\begin{equation}
\mathcal{F}(\eta) = \frac{e^{-\gamma_E\eta}}{\Gamma(1+\eta)}\left(\frac{cQ}{\mu_S}\right)^{\!\eta}.
\end{equation}
Define $\Phi = \partial_\eta \ln\mathcal{F} = L_c - \gamma_E - \psi(1+\eta)$ where $L_c = \ln(cQ/\mu_S)$ and $\psi(x) = \Gamma'(x)/\Gamma(x)$. Then $\partial_\eta$ acting on $\mathcal{F}$ produces $\Phi\,\mathcal{F}$, and $\partial_\eta + \ln(Q\mu_S/\mu_J^2)$ produces $(\Phi + \ln(Q\mu_S/\mu_J^2))\,\mathcal{F} = (\ln(cQ^2/\mu_J^2) - \gamma_E - \psi(1+\eta))\,\mathcal{F}$. The resummed cumulant is
\begin{equation}
\label{eq:app-R-final}
R(c) = \Pi(\mu_H,\mu_J,\mu_S)\;h(L_H)\;\tilde{j}_q\!\left(\partial_\eta + \ln\frac{Q\mu_S}{\mu_J^2}\right)^{\!2}\,\tilde{j}_g\!\left(\partial_\eta + \ln\frac{Q\mu_S}{\mu_J^2}\right)\,\tilde{s}(\partial_\eta)\;\mathcal{F}(\eta)\,,
\end{equation}
where $\partial_\eta$ acts on $\mathcal{F}(\eta)$ and $L_H = \ln(Q^2/\mu_H^2)$.


\section{Resummation Formula}
\label{app:NLL}

This appendix collects the complete formulas for NLL resummation of the C-parameter shoulder.

The resummed cumulant takes the general form
\begin{equation}
\label{eq:R-general-app}
R(c) = \Pi(\mu_H, \mu_J, \mu_S)\; h(L_H)\; \tilde{j}_q\big(\partial_\eta + \ln\frac{Q\mu_S}{\mu_J^2}\big)^2\, \tilde{j}_g\big(\partial_\eta + \ln\frac{Q\mu_S}{\mu_J^2}\big)\, \tilde{s}(\partial_\eta)\; \mathcal{F}(\eta)\,,
\end{equation}
where the evolution kernel is
\begin{align}
\Pi(\mu_H, \mu_J, \mu_S) &= \exp\Big[2\mathcal{C}\big(S(\mu_H, \mu_J) + S(\mu_S, \mu_J)\big) - 2A_{\gamma_H}(\mu_H,\mu_J) + 2A_{\gamma_S}(\mu_S,\mu_J)\Big] \nn\\
&\quad \times \left(\frac{Q^2}{\mu_H^2}\right)^{-\mathcal{C}\,A_\Gamma(\mu_H, \mu_J)},
\end{align}
$\mathcal{C} = 2\CF + \CA$ is the total cusp color factor, the hard logarithm is $L_H = \ln(Q^2/\mu_H^2)$, the base function is
\begin{equation}
\label{eq:base-F}
\mathcal{F}(\eta) = \frac{e^{-\gamma_E\eta}}{\Gamma(1+\eta)}\left(\frac{cQ}{\mu_S}\right)^{\!\eta},
\end{equation}
and the Sudakov exponent is $\eta = 2\mathcal{C}\, A_\Gamma(\mu_J, \mu_S)$. The following subsections provide explicit expressions for all ingredients.

\subsection{Perturbative constants}
The beta function and cusp anomalous dimension coefficients are:
\begin{align}
\beta_0 &= \frac{11\CA - 4\TF\nf}{3}\,, &
\beta_1 &= \frac{34\CA^2}{3} - \frac{20\CA\TF\nf}{3} - 4\CF\TF\nf\,, \\[0.3em]
\Gamma_0 &= 4\,, &
\Gamma_1 &= 4\left[\left(\frac{67}{9} - \frac{\pi^2}{3}\right)\CA - \frac{20}{9}\TF\nf\right].
\end{align}

The one-loop non-cusp anomalous dimensions for the C-shoulder jet functions (which include the $\ln 3$ shift from the azimuthal average), soft function, and hard function are:
\begin{align}
\gamma_{J,q}^{C,(0)} &= -3\CF - 4\CF\ln 3\,, \\[0.3em]
\gamma_{J,g}^{C,(0)} &= -\beta_0 - 4\CA\ln 3\,, \\[0.3em]
\gamma_S^{(0)} &= 2\mathcal{C}\ln 3\,, \\[0.3em]
\gamma_H^{(0)} &= -6\CF - \beta_0 - 2\mathcal{C}\ln 3\,.
\end{align}
These satisfy RG consistency: $\gamma_H = 2\gamma_{J,q}^C + \gamma_{J,g}^C + \gamma_S$. The $\ln 3$ terms encode the Mercedes geometry ($s_{ij} = Q^2/3$, $n_i \cdot n_j = 3$).

The combined jet anomalous dimension is
\begin{equation}
\gamma_J^{C,(0)} \equiv 2\gamma_{J,q}^{C,(0)} + \gamma_{J,g}^{C,(0)} = -6\CF - \beta_0 - 4\mathcal{C}\ln 3\,.
\end{equation}

The one-loop matching constants are set to zero at NLL. The hard and soft constants are:
\begin{align}
c_H^1 &= \CF\left[-\frac{65}{4} + \frac{3\pi^2}{2} - \frac{21}{8}\ln 3 - 10\ln 2\ln 3 + 3\ln^2 3 + 10\,{\rm Li}_2\!\left(\tfrac{1}{3}\right)\right] \nn\\
&\quad + \CA\left[\frac{3}{4} + \frac{5\pi^2}{4} + \frac{3}{8}\ln 3 + \ln 2\ln 3 - \frac{3}{2}\ln^2 3 - {\rm Li}_2\!\left(\tfrac{1}{3}\right)\right], \\[0.3em]
c_S^1 &= \mathcal{C}\left(-\frac{2\pi^2}{3} + 4\ln^2 2 - \frac{1}{3}\ln 27\ln 48 + 2\,{\rm Li}_2\!\left(-\tfrac{1}{3}\right)\right).
\end{align}
The inclusive jet matching constants in Laplace space are:
\begin{equation}
c_J^q = \CF\left(7 - \frac{2\pi^2}{3}\right), \qquad c_J^g = \CA\left(\frac{67}{9} - \frac{2\pi^2}{3}\right) - \frac{20}{9}\TF\nf\,.
\end{equation}
The C-shoulder jet matching constants are shifted from the inclusive jet constants:
\begin{equation}
c_J^{C,i} = c_J^i - \gamma_{J,i}\ln 3 + \frac{C_i\Gamma_0}{2}\left(\ln^2 3 + \frac{\pi^2}{3}\right).
\end{equation}
The combined jet matching constant is $c_J^C \equiv 2c_J^{C,q} + c_J^{C,g}$.

\subsection{Matching functions}
The Laplace-space matching functions at one loop are:
\begin{equation}
h(L_H) = 1 + \frac{\as}{4\pi}\left[-\mathcal{C}\frac{\Gamma_0}{4}L_H^2 - \gamma_H^{(0)} L_H + c_H^1\right]
\end{equation}
where $L_H = \ln(Q^2/\mu_H^2)$.

The jet matching function for parton $i \in \{q,g\}$ with Casimir $C_i$ ($C_q = \CF$, $C_g = \CA$) is:
\begin{equation}
\tilde{j}_i^C(L) = 1 + \frac{\as}{4\pi}\left[C_i\frac{\Gamma_0}{2}L^2 + \gamma_{J,i}^{C,(0)} L + c_J^{C,i}\right]
\end{equation}
where the jet functions are evaluated at $L \to L + \ln(Q\mu_S/\mu_J^2)$.

The soft matching function is:
\begin{equation}
\tilde{s}(L) = 1 + \frac{\as}{4\pi}\left[-\mathcal{C}\Gamma_0 L^2 + 2\gamma_S^{(0)} L + c_S^1\right]
\end{equation}
where $L = \ln(\mu_S/Q)$.

\subsection{Evolution}
The strong coupling at scale $\mu$ in terms of the coupling at scale $Q$ is given at two loops by
\begin{equation}
\as(\mu) = \as(Q)\left[X + \as(Q)\frac{\beta_1}{4\pi\beta_0}\ln X\right]^{-1}, \qquad X \equiv 1 + \frac{\as(Q)}{2\pi}\beta_0\ln\frac{\mu}{Q}\,.
\end{equation}

With $r = \as(\mu)/\as(\nu)$, the Sudakov integral at NLL is:
\begin{equation}
\label{eq:S-NLL}
S(\nu,\mu) = \frac{\Gamma_0\pi}{\beta_0^2\as(\nu)}\left(1 - \frac{1}{r} - \ln r\right) + \frac{\Gamma_0}{4\beta_0^2}\left[\left(\frac{\Gamma_1}{\Gamma_0} - \frac{\beta_1}{\beta_0}\right)(1 - r + \ln r) + \frac{\beta_1}{2\beta_0}\ln^2 r\right].
\end{equation}

The cusp evolution integral at NLL is:
\begin{equation}
A_\Gamma(\nu,\mu) = \frac{\Gamma_0}{2\beta_0}\ln r + \frac{\Gamma_0\as(\nu)}{8\pi\beta_0}\left(\frac{\beta_1}{\beta_0} - \frac{\Gamma_1}{\Gamma_0}\right)(1 - r)\,.
\end{equation}

The non-cusp evolution integral at NLL is:
\begin{equation}
\label{eq:Agamma-NLL}
A_{\gamma_X}(\nu,\mu) = \frac{\gamma_X^{(0)}}{2\beta_0}\ln r\,.
\end{equation}

The evolution kernel is given by:
\begin{equation}
\label{eq:Pi-evolution}
\Pi(\mu_H, \mu_J, \mu_S) = \exp\Big[2\mathcal{C}\big(S(\mu_H, \mu_J) + S(\mu_S, \mu_J)\big) - 2A_{\gamma_H}(\mu_H,\mu_J) + 2A_{\gamma_S}(\mu_S,\mu_J)\Big] \left(\frac{Q^2}{\mu_H^2}\right)^{\!-\mathcal{C}\,A_\Gamma(\mu_H, \mu_J)},
\end{equation}
and the Sudakov exponent is
\begin{equation}
\eta = 2\mathcal{C}\, A_\Gamma(\mu_J, \mu_S)\,.
\end{equation}

\subsection{Kernel coefficients}
\label{sec:kernel-coefficients}

For practical implementation, we expand the product of matching functions in Eq.~\eqref{eq:R-general-app} and act with the derivatives on $\mathcal{F}$. The base function is
\begin{equation}
\mathcal{F}(\eta) = \frac{e^{-\gamma_E\eta}}{\Gamma(1+\eta)}\left(\frac{cQ}{\mu_S}\right)^{\!\eta},
\end{equation}
where $\eta = 2\mathcal{C}\, A_\Gamma(\mu_J, \mu_S)$. At NLL the cumulant can be written as
\begin{equation}
\label{eq:R-knFn}
R(c) = \Pi(\mu_H, \mu_J, \mu_S)\left(k_0 + k_1 F_1 + k_2 F_2\right)\mathcal{F}\,,
\end{equation}
where $\Pi$ is given in Eq.~\eqref{eq:Pi-evolution}. The derivative functions $F_n \equiv (\partial_\eta^n\mathcal{F})/\mathcal{F}$ are
\begin{equation}
F_1 = \partial_\eta\ln\mathcal{F} = \ln\frac{cQ}{\mu_S} - \gamma_E - \psi(1+\eta)\,,
\end{equation}
and
\begin{equation}
F_2 = F_1^2 - \psi_1\,,
\end{equation}
where $\psi_1 = d\psi/dx|_{x=1+\eta}$.

To obtain the kernel coefficients $k_n$, we compute the product $h\cdot \tilde{j}_q^2\cdot \tilde{j}_g\cdot \tilde{s}$ using the matching functions from the previous subsection, expand in powers of $L = \ln(\mu_S/Q)$, and replace $L^n \to F_n$.
Defining $L_H = \ln(Q^2/\mu_H^2)$ and $L_R = \ln(Q\mu_S/\mu_J^2)$, and using $\gamma_J^{C,(0)}$ and $c_J^C$ from Section~D.1, the NLL kernel coefficients are:
\begin{align}
\label{eq:k1-coeffs}
k_0 &= \frac{\as}{4\pi}\left[c_H^1 + c_J^C + c_S^1 - \frac{\mathcal{C}\Gamma_0}{4}L_H^2 + \frac{\mathcal{C}\Gamma_0}{2}L_R^2 - \gamma_H^{(0)} L_H + \gamma_J^{C,(0)} L_R\right], \nn\\[0.5em]
k_1 &= \frac{\as}{4\pi}\left[\mathcal{C}\Gamma_0 L_R + \gamma_J^{C,(0)} + 2\gamma_S^{(0)}\right], \\[0.5em]
k_2 &= -\frac{\as}{4\pi}\,\frac{\mathcal{C}\Gamma_0}{2}\,. \nn
\end{align}
At NLL accuracy, the matching constants are set to zero: $c_H^1 = c_J^C = c_S^1 = 0$.

\subsection{Fixed-order expansion by color structure}
\label{sec:fixed-order-color}

The resummed cumulant Eq.~\eqref{eq:R-knFn} can be expanded at equal scales $\mu_H = \mu_J = \mu_S = Q$ to obtain fixed-order predictions. The $\order{\as}$ cumulant with $L = \ln c$ is
\begin{align}
R_1(L) &= \frac{\as}{2\pi}\left[-\frac{\mathcal{C}\Gamma_0}{4} L^2 + \frac{1}{2}\big(\gamma_J^{C,(0)} + 2\gamma_S^{(0)}\big) L + \frac{c_{\rm tot}}{2}\right] \nn\\[0.3em]
&= \frac{\as}{2\pi}\left\{\CF\left[-2L^2 - 3L\right] + \CA\left[-L^2 - \frac{11}{6}L\right] + \TF\nf\left[\frac{2}{3}L\right]\right\},
\label{eq:R1-color}
\end{align}
where $c_{\rm tot} = c_H^1 + 2c_J^{C,q} + c_J^{C,g} + c_S^1$ is the sum of matching constants (which vanish at NLL). The coefficient of $L^2$ is $-\mathcal{C}\Gamma_0/4 = -\mathcal{C} = -(2\CF + \CA)$, reproducing the Catani--Webber result $A_2 = 2\CF + \CA$. The single-log coefficient $(\gamma_J^{C,(0)} + 2\gamma_S^{(0)})/2 = -3\CF - \beta_0/2$ equals $-B_1$ from Eq.~\eqref{eq:B1-coeff}; remarkably, the $\ln 3$ terms from the Mercedes geometry cancel in this combination. Combined with the leading $\alpha_s A_{\rm Born}(3/4)$ factor, the LO singular cross section agrees with Eq.~\eqref{eq:SCET-full-expansion}.



\end{document}